\documentclass[11pt]{article}
\usepackage{epsfig}
\usepackage{graphicx, subfigure}
\usepackage{amsmath}
\usepackage{amsfonts}
\usepackage{latexsym}

\newtheorem{theorem}{Theorem}
\newtheorem{lemma}{Lemma}
\newtheorem{corollary}{Corollary}
\newtheorem{proposition}{Proposition}
\newtheorem{definition}{Definition}

\newtheorem{remark}{{\bf Remark}} 

\newlength{\dinwidth}
\newlength{\dinmargin}
\def\beqs{\begin{displaymath}}
\def\eeqs{\end{displaymath}}
\def\beqn{\begin{eqnarray}}
\def\eeqn{\end{eqnarray}}

\def\lef{\gamma}
\def\i{{\rm i}}

\def\Tq{{\bf{T_q}}}
\def\Tdoubles{{\bf T}}
\def\Id{{\bf{1}}}

\def\lb{\bar{\l}}
\def\Qbar{\bar{Q}}
\def\Pbar{\bar{P}}
\def \q {{\bf {q}}}
\def\Wq{W_\q}

\def\M{{\cal H}}
\def \F{{\cal F}}

\def\calT{{\cal T}}
\def\calW{{\cal W}}
\def\calV{{\cal V}}

\def\c{{\cal C}}

\def\cp{\mathbb {CP}^1}
\def\l{\lambda}
\def\C{\mathbb {C}}

\def\R{\mathbb {R}}

\def\F{{\cal F}}
\def\d{\partial}
\def\hsp{\hspace{0.5cm}}
\def\imb{\mathrm{Im}\mathbb B}
\def\B{\mathbb B}

\def \surf {{\cal L}}
\def\PsiSB{{\Psi_{\iOmega \iB}}}

\def\iE{{\scriptscriptstyle{E}}}
\def\iT{{\scriptscriptstyle{T}}}

\def\iP{{\scriptscriptstyle{P}}}
\def\iQ{{\scriptscriptstyle{Q}}}
\def\iOmega{{\scriptscriptstyle{\Omega}}}
\def\iB{{\scriptscriptstyle{B}}}
\def\io{{\scriptscriptstyle{0}}}
\def\ione{{\scriptscriptstyle{1}}}
\def\itwo{{\scriptscriptstyle{2}}}
\def\ithree{{\scriptscriptstyle{3}}}
\def\ifour{{\scriptscriptstyle{4}}}
\def\iW{{\scriptscriptstyle{W}}}

\def\iI{{\scriptscriptstyle{I}}}

\renewcommand{\thefootnote}{}

\begin{document}

\title{Riemann-Hilbert problem associated to Frobenius manifold structures on  Hurwitz spaces: irregular singularity}
\author{Vasilisa Shramchenko}        
\date{}   
\maketitle

{\bf Abstract.}  Solutions to the Riemann-Hilbert problems with irregular singularities naturally associated to semisimple Frobenius manifold structures on Hurwitz spaces (moduli spaces of meromorphic functions on Riemann surfaces) are constructed. The solutions are given in terms of meromorphic bidifferentials defined on the underlying Riemann surface.
The relationship between  different classes of Frobenius manifolds structures on Hurwitz spaces
 (real doubles, deformations) is described at the level of the corresponding Riemann-Hilbert problems.


\footnotetext{This work was supported in part by the Humboldt Foundation and the Engineering and Physical Sciences Research Council.}

\section{Introduction}

In its original formulation,  the Riemann-Hilbert  (Hilbert's 21st) problem  is the problem of exis\-tence of a Fuchsian system of linear ordinary differential equations on the Riemann sphere having given singularities and monodromy data. 

Another problem is to find a solution to that system. This problem can be reformulated in a general form as  the problem of reconstructing an analytic function in the complex plane from jump conditions across some curves. The latter problem is also traditionally referred to as the Riemann-Hilbert (factorization) problem; it is this meaning of the term that we adopt here.
Problems of this type play an important role in the theory of integrable systems, see the review \cite{Its}.

A solution to the Riemann-Hilbert problem can be rarely expressed in terms of known special functions. There are, however, examples of monodromy data for which explicit solutions have been found. For instance, in \cite{DimaRH} the Riemann-Hilbert  problem corresponding to a Fuchsian system was solved for quasi-permutation monodromy matrices. The matrix solution was found in terms of a generalization of the Szeg\"o kernel on a Riemann surface. In the matrix dimension two, the explicitly solvable Riemann-Hilbert problems with off-diagonal monodromies are used in the theory of random matrices as asymptotic Riemann-Hilbert problems in the study of the large $N$ limit of Hermitian matrix models \cite{DIZ}.

In this paper we present solutions to another class of Riemann-Hilbert problems. The problems studied here arise in the theory of Frobenius manifolds, which were introduced \cite{2D} as a geometric formulation of the structure of the Witten-Dijkgraaf-Verlinde-Verlinde (WDVV) equations -- the structure which also appears in singularity theory, in the theory of Gromov-Witten invariants of projective varieties and in other branches of mathematics, see \cite{2D, DubrovinPainleve, Manin}. 

We consider Frobenius ma\-ni\-fold structures on Hurwitz spaces, i.e., on the moduli spaces of pairs -- a Riemann surface and a meromorphic function on the surface. There exist three classes of Frobenius manifold structures on Hurwitz spaces. The first class was found in \cite{2D}. The other two classes \cite{doubles, deform} are the so-called  real doubles   and  deformations  of the manifolds from \cite{2D}. All these Frobenius structures are {\it semisimple} (i.e., such that the algebras defined in the tangent spaces to the manifolds contain no nilpotents). 

Our motivation to study the Riemann-Hilbert problems related to Frobenius manifolds is based on the fact that a semisimple Frobenius manifold can be reconstructed \cite{DubrovinPainleve} from a fundamental solution to the associated Riemann-Hilbert problem. 
Therefore the relationships between different Frobenius manifold structures on Hurwitz spaces may potentially be revealed by studying the corresponding Riemann-Hilbert problems. Furthermore, such
relationships could allow to extend the notions of  real doubles and deformations to an arbitrary semisimple Frobenius manifold. 
In this paper we make a step in this direction by finding transformations relating solutions to the Riemann-Hilbert problems corresponding to the three classes of Frobenius manifold structures on Hurwitz spaces mentioned above.

There are two Riemann-Hilbert problems -- a Fuchsian and a non-Fuchsian one -- associated to each semisimple Frobenius manifold. The problems are dual to each other, i.e., related by a formal Laplace transform \cite{2D}. In this paper we study the non-Fuchsian problem, which is
associated to a Frobenius manifold in the following way. 
Given a point on a semisimple Frobenius manifold $\F$ of dimension $n$, one can construct the following non-Fuchsian matrix ordinary differential equation on $\cp$ with rational coefficients:
\beqn
\frac{\d \Psi}{\d z} = (U+\frac{1}{z}V)\Psi .
\label{RHz} 
\eeqn
Here $\Psi$ is an $n\times n$ matrix-valued function on $\cp\times \F;$ and
$U$ is the diagonal matrix $U={\rm diag}\{\l_1,\dots,\l_n\}$, where $\{\l_i\}$ is a distinguished set of coordinates on the Frobenius manifold, the  {\it canonical coordinates} (see Section \ref{sect_HF}). The skew-symmetric matrix $V$ is determined by the Frobenius manifold as follows: 
\beqn
V:=[\Gamma,U],
\label{V}
\eeqn
entries of the symmetric matrix $\Gamma$ are given by the {\it rotation coefficients} $\beta_{ij}$ (defined below by (\ref{rot-def})) of the flat metric on the Frobenius manifold: $\Gamma_{ij}=\beta_{ij}$ if $i\neq j$ and $\Gamma_{ii}=0.$

Equation (\ref{RHz})  has a regular singularity at the origin and an irregular singularity at the point at infinity. The monodromy data of this equation is called the monodromy data of the cor\-res\-pon\-ding Frobenius manifold; it was used for classifying all semisimple Frobenius manifolds \cite{DubrovinPainleve}. The monodromy data includes the Stokes matrix at irregular singularity and the behaviour of a solution at singular points. 

The  Frobenius manifold axioms imply the {\it isomonodromy condition}: the independence of the 
monodromy at the origin and the Stokes matrix
of equation (\ref{RHz}) of the point of the  Frobenius manifold. This condition is equivalent to the compatibility of equation (\ref{RHz}) with the following system defining the dependence of the function $\Psi$ on the canonical coordinates on the manifold:
\beqn
\frac{\d \Psi}{\d \l_i} = (zE_i - [E_i,\Gamma])\Psi , \hsp i=1,\dots, n.
\label{RHi}
\eeqn
Here, $E_i$ is the matrix having only one non-vanishing entry, which is equal to one and stands on the $i$th place on the diagonal. 
 In the matrix dimension $3$, the compatibility condition of the system (\ref{RHz}), (\ref{RHi}) is equivalent to the Painlev\'e-VI equation with the coefficients $(1/8,-1/8,1/8,3/8),$ see \cite{2D}.
 
 The fundamental solution to system (\ref{RHz}), (\ref{RHi})  contains the complete information about the underlying Frobenius manifold structure \cite{DubrovinPainleve}. 

 The problem of reconstructing the solution to equation (\ref{RHz}) from its monodromy data is the Riemann-Hilbert problem associated to the Frobenius manifold.

The description of the Frobenius manifold structures on Hurwitz spaces includes representations of the matrices $V$ and $\Gamma$ in terms of various meromorphic bidifferentials defined on the Riemann surface.
The main example of such a bidifferential is given by  the {\it canonical nor\-ma\-lized bidifferential W(P,Q)}, which is symmetric and has vanishing $a$-periods and a second order pole on the diagonal. Other bidifferenitals used in the construction include the Schiffer and Bergman kernels and a deformation of the bidifferential W. 

The purpose of this paper is to solve the system (\ref{RHz}), (\ref{RHi}) for  Frobenius manifold structures on Hurwitz spaces. 
The solutions are given in terms of the bidifferentials mentioned above. 
For instance, in the case corresponding to the $n$-dimensional Frobenius structures from \cite{2D} on Hurwitz spaces of the pairs $(\surf,\l)$ the solution is represented in terms of the bidifferential $W.$ It is given by an $n\times n$ matrix with the entries:
\beqn
\left(\Psi(z)\right)_{ij}:=\frac{1}{2 \i \sqrt{\pi}} \frac{1}{\sqrt{z}} \int_{\c_j} {\rm e}^{z\,\l(Q)}W(Q,P_i).
\label{introPsi}
\eeqn
Here, $z\in\cp;$ the set  $\{P_i\}_{i=1}^n$ is the set of simple critical points of the meromorphic function $\l$ on the surface $\surf;$  and $Q$ stands for a point on the surface; the bidifferential $W$ is evaluated at  $P_i$ with respect to a local parameter defined by the function $\l.$
The integration paths $\{\c_j\}_{j=1}^n$ are given by contours on the Riemann surface which go from a pole of the meromorphic function $\l$ to another pole  through a neighbourhood of one of the critical points $\{P_j\}_{j=1}^n$ in a prescribed way.  
The direction in which the contours approach poles of the meromorphic function depends on ${\rm arg}\{z\}.$
 Thus, the monodromy of $\Psi(z)$  is related to the monodromy in the appropriate space of contours on the Riemann surface. 

As an example we consider Frobenius manifold structures on the Hurwitz space of degree $2$ meromorphic fun\-ctions on Riemann surfaces of genus $g.$ The dimension of this space and the associated Frobenius structure is $n=2g+2;$ the local coordinates on the Hurwitz space (and the canonical coordinates on the corresponding Frobenius structure) are given by the critical values of the meromorphic functions. 
Under a certain ordering of the critical points,  the  Stokes matrix  of the corresponding solution  (\ref{introPsi}) is given by the $n\times n$ lower triangular matrix $S$ with the entries:
\beqs
\begin{array}{ll}
s_{ii}=1;&\qquad\\
s_{ij}=(-1)^{i+j}2, \hsp &i<j;\\
s_{ij}=0, \hsp &i>j.
\end{array}
\eeqs

The solutions to the Riemann-Hilbert problems associated to the deformations and real doubles of Dubrovin's \cite{2D} Frobenius structures  on Hurwitz spaces turn out to be related to the  solution (\ref{introPsi}) by a Schlesinger transformation, also known as {\it dressing transformation}. 
 Namely, the solution corresponding to the deformations, we denote it by $\Psi_\q,$ 
is obtained from the  function $\Psi$  given by (\ref{introPsi}) as follows:
\beqs
\Psi_\q (z) = \left(\Id -  \frac{1}{z} \Tq \right) \Psi (z) ,
\eeqs
where $\Id$ denotes the identity matrix; and $\Tq$ is a matrix function on the Frobenius manifold independent of $z,$ such that ${\bf T}^2_{\bf q}$ vanishes.
The Stokes matrices of the Frobenius structures on Hurwitz spaces from \cite{2D} thus coincide with the Stokes matrices of their deformations. 

In the case of the real doubles, the solution to the Riemann-Hilbert problem is constructed in terms of the Schiffer $\Omega$ and Bergman $B$ kernels on the Riemann surface; we denote the solution by $\PsiSB.$ Its matrix dimension is twice as large as the dimension of the matrix $\Psi(z)$ given by (\ref{introPsi}). The solution $\PsiSB$  is obtained by a Schlesinger transformation 
from a block-diagonal matrix whose blocks are built from the function $\Psi(z)$  as follows:  
\beqs
\PsiSB (z) = \left(\Id -  \frac{1}{z} \Tdoubles \right) \left(\begin{array}{cc}\Psi(z) & 0 \\0 &\overline{\Psi(\bar{z})} \end{array}\right) .
\eeqs
Here, $\Tdoubles$ is a matrix-valued function on the Frobenius manifold independent of $z$ (and such that $\Tdoubles^2=0$). Thus, the Stokes matrix  for the real doubles of a Frobenius manifold $\F$ is  the block-diagonal matrix  whose blocks are given by the Stokes matrix of the manifold $\F.$

The above Schlesinger transformations are constructed  for solutions related to the Hurwitz spaces: the expressions for the matrices $\Tq$ and $\Tdoubles$ are written in terms of differentials defined on the underlying Riemann surface. A representation of these transformations in terms  universal to the  Frobenius structure could lead to construction of a real double and a deformation of an arbitrary semisimple Frobenius manifold.

The paper is organized in the following way. Section \ref{sect_RH} is devoted to a general discussion of a solution to the system (\ref{RHz}), (\ref{RHi}) and the Riemann-Hilbert problem. Section \ref{sect_Hurwitz} contains the necessary  facts from the theory of Hurwitz spaces and Frobenius manifolds. Section \ref{sect_solution} is the main section of the paper -- it contains the solutions of the Riemann-Hilbert problems associated to the Frobenius manifold structures on  Hurwitz spaces. In this section we also describe the relationship between the Riemann-Hilbert problems corresponding to the Frobenius structures on Hurwitz spaces from \cite{2D}, their real doubles and deformations. In Section \ref{sect_examples} we explicitly compute the Stokes matrices in the case of the Hurwitz spaces of functions with simple poles.

\section{The Riemann-Hilbert problem}

\label{sect_RH}

Generally speaking,  a Riemann-Hilbert problem is the problem of reconstructing a matrix function with a singularity of a given type and a given discontinuity. We study a problem of this kind associated to the linear system (\ref{RHz}), (\ref{RHi}).  We first look at the general behaviour of a solution to the system and then formulate the corresponding Riemann-Hilbert problem. 
This section collects the results of \cite{2D, DubrovinPainleve, Wasow} which are used in the sequel.

\subsection{Behaviour of the solution and monodromy data}
\label{sect_monodromy}

A  solution to equation (\ref{RHz}) has an irregular singularity of Poincar\'e rank $1$ at the point at infinity. There exists a formal solution $\tilde{\Psi}$ of the form:
\beqn
\tilde{\Psi}(z) = \left(  \Id  +  \frac{\Gamma_1}{z} + \frac{\Gamma_2}{z^2}+ \dots \right) {\rm e}^{zU} , 
\label{formal}
\eeqn
where $U$ is the diagonal matrix from (\ref{RHz}), $\Gamma_i$ are matrix functions of the coordinates $\{\l_k\}$ with $\Gamma_1=\Gamma$ being the matrix of rotation coefficients from (\ref{V}).  There are sectors in the $z$-plane where fundamental solutions $\Psi(z)$ to equation (\ref{RHz}) with the asymptotic behaviour 
\beqn
\label{atinf}
\Psi(z)\sim\tilde{\Psi}(z) \hsp \mbox{as} \qquad |z| \to \infty
\eeqn
 exist. To describe these sectors we start with the following definitions.
\begin{definition}
\label{def-admissible}
{\rm A line $l$ going through the origin in the complex $z$-plane is called {\it admissible} for equation (\ref{RHz}) if $\;{\rm Re \;} \{z(\l_i-\l_j)\} \neq 0$ for nonzero $z\in l$ and for any $i\neq j . $}
\end{definition}
The non-admissible directions of equation (\ref{RHz}) are given by
the so-called Stokes rays. They are determined by the configuration of the eigenvalues of the matrix $U$ as follows. 
\begin{definition}
{\rm The rays $r_{ij} = - r_{ji}$ defined for distinct indices $i$ and $j$ by
\beqn
\label{rays}
r_{ij}:= \{ z \mid {\rm{Re}} \left[ z ( \l_i - \l_j ) \right]=0 ,\;\; {\rm Im} [  z ( \l_i - \l_j ) ] < 0  \} 
\eeqn
and oriented from the origin are called the {\it Stokes rays} of equation (\ref{RHz}). }
\end{definition}

\begin{figure}[htb]
\centering
\subfigure{\includegraphics[width=6cm]{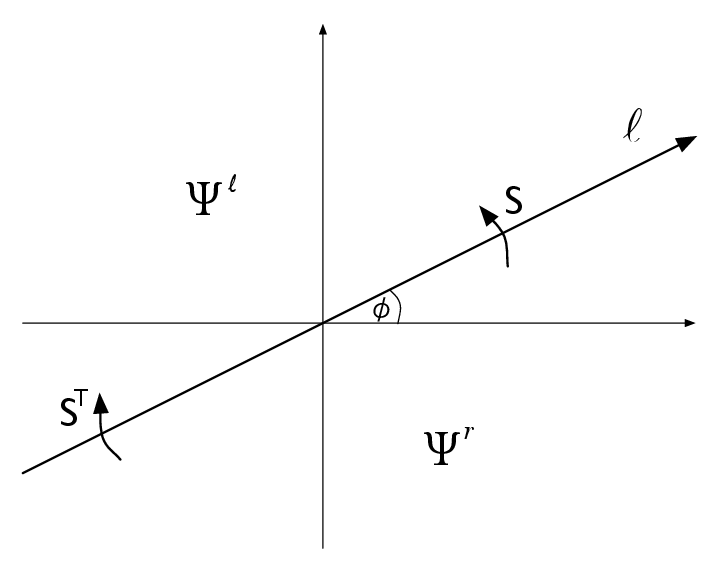}}
\caption{$z$-plane.}
\label{fig_zplane}
\end{figure}

Let us fix an oriented admissible line $l.$ It divides the $z$-plane into the left $\Pi^l$ and right $\Pi^r$ half-planes. Let $\phi$ be the angle between the  line $l$ and the real axis. Consider the domains given by  sectorial neighbourhoods of the half-planes: 
\beqn
\label{rl}
\Pi^l_\varepsilon = \{z \mid \phi - \varepsilon <{\rm arg} (z)< \pi +\phi +\varepsilon \} , \;\;\;\; 
\Pi^r_\varepsilon = \{z \mid \phi - \pi - \varepsilon <{\rm arg} (z)< \phi + \varepsilon \} ,
\eeqn
where $\varepsilon>0$ is sufficiently small, i.e., such that $\Pi^l_\varepsilon$ (respectively $\Pi^r_\varepsilon$) contains only the Stokes rays contained in $\Pi^l$ (respectively $\Pi^r$). In such neighbourhoods $\Pi^r_\varepsilon$ and $\Pi^l_\varepsilon$ of any half-plane there exist  \cite{Wasow}  solutions to equation (\ref{RHz}) with the  asymptotics as $|z|\to \infty $ given by the formal solution (\ref{formal}). We shall denote the respective solutions by $\Psi^{r}$ and $\Psi^l,$ see Figure \ref{fig_zplane}.

The intersection of the domains $\Pi^r_\varepsilon$ and $\Pi^l_\varepsilon$ consists of two sectors;
each of the sectors contains a part of the separating line $l.$
Let us orient these parts in the direction from the origin and denote the obtained rays by $l_+$ and $l_-$:  the orientation of $l_+$ coincides with that of the line $l$. There are  two solutions $\Psi^r$ and $\Psi^l$ to equation (\ref{RHz}) defined in each of the sectors of the intersection $\Pi_\varepsilon^r \bigcap \Pi_\varepsilon^l.$ These solutions are therefore related in each sector by a right multiplication with a $z$-independent matrix. Let us denote the matrix relating the solutions in the sector containing $l_+$ by $S:$
\beqn
\label{SM}
\Psi^l(z) = \Psi^r(z)S.
\eeqn
\begin{definition}
{\rm The matrix $S$ in (\ref{SM}) is called the {\it Stokes matrix} of equation (\ref{RHz}).}
\end{definition}
The matrix relating the solutions in the sector containing $l_-$ equals $S^\iT$ as a corollary of the skew-symmetry of the matrix $V$ from equation (\ref{RHz}): $\Psi^l(z) = \Psi^r(z)S^\iT, \;\; z\in l_-$. 

Thus the matrix 
\beqs
M_\infty:=S(S^{\iT})^{-1}
\eeqs
gives the monodromy of the solution with respect to analytic continuation around the point $z=\infty$ counterclockwise starting in $\Pi^r.$

The Stokes matrix has the following structure: its diagonal entries equal one, and an off-diagonal entry $S_{ij}$ vanishes if the corresponding Stokes ray $r_{ij}$ (\ref{rays}) belongs to the right half-plane $\Pi^r.$ This can be seen from the asymptotics (\ref{formal}), (\ref{atinf}), which implies ${\rm e}^{zU}S{\rm e}^{-zU} \to \Id$ as $z\to\infty$ along the ray $l_+.$ Therefore, $S_{ij}=0$ if ${\rm Re}\{z(\l_i-\l_j)\}>0$ or, equivalently, if $r_{ij}\in\Pi^r.$

A domain of existence of  a solution with asymptotics (\ref{formal}), (\ref{atinf}) can be broadened \cite{BJL, Wasow}. Namely,   the parameter $\varepsilon$  in the definition of $\Pi^l_\varepsilon$  can be increased until one of the rays $\{z \mid {\rm arg} (z)=\pi +\phi +\varepsilon \}$ or $\{z \mid  {\rm arg} (z)=\phi - \varepsilon \}$  bordering the domain meets a  Stokes ray of the equation. 
The analytic continuation of the solution defined in $\Pi_\varepsilon^l$  into a neighbourhood of $z=\infty$ beyond this Stokes ray no longer has the asymptotic behaviour (\ref{atinf}). The analogous holds for the analytic continuation of $\Psi^r$ defined in  $\Pi^r_\varepsilon.$

At the origin a solution to equation (\ref{RHz}) has  a regular singularity.  The expansion of $\Psi^r$ in a neighbourhood of $z=0$ intersected with $\Pi^r$ has the form
\beqn
\Psi^r(z) \simeq  G(z)  z^\mu z^R C_\io^r , \qquad z\sim 0,
\label{Psi_r_at0}
\eeqn
where $C_\io^r$ is a constant matrix; $\mu = {\rm diag}\{\mu_1,\dots,\mu_n\}$ is the diagonalization of the skew-symmetric matrix $V$ from (\ref{RHz}); $G(z)$ is holomorphic at $z=0$ and such that $G(0)$ is non-degenerate and sa\-tis\-fies $ G(0)^{-1}V G(0) =\mu;$ 
 the matrix $R$ appears in the {\it resonant} case, i.e., when some eigenvalues of the matrix $V$ differ by an integer: the matrix $R$ satisfies $R_{ij} \neq 0$ only if $\mu_i-\mu_j=k$ for a positive integer $k.$

A solution of the form (\ref{Psi_r_at0}) can be analytically continued into a disc neighbourhood of $z=0$ with a branch cut ending at the origin.  Thus  the monodromy matrix of $\Psi^r$ at the origin has the form $M = (C_\io^r)^{-1} {\rm e}^{2 \pi \i \mu} {\rm e}^{2\pi \i R}C_\io^r.$ 

The two monodromy matrices at singular points of the equation satisfy $M_\infty M=\Id,$ which is equivalent to the relation $S^\iT S^{-1} = (C_\io^r)^{-1} {\rm e}^{2 \pi \i \mu} {\rm e}^{2\pi \i R}C^r_\io.$

The monodromy matrices $M_\infty$ and $M$ as well as the Stokes matrix $S$ and the matrices $C_\io^r,$ $\mu$ and $R$ satisfy the isomonodromy condition, i.e., they do not depend \cite{2D} on the point of the Frobenius manifold. This is a consequence of the compatibility of equations (\ref{RHz}) and (\ref{RHi}), which follows from the axioms of the Frobenius manifold structure. 

Reconstruction of the matrix functions $\Psi^r$ and $\Psi^l$ from the given 
monodromy data including the Stokes  matrix $S$ and the asymptotics  (\ref{formal}), (\ref{atinf}) and (\ref{Psi_r_at0}) near singular points amounts to solving the following Riemann-Hilbert problem.

\subsection {Formulation of the Riemann-Hilbert problem}
\label{sect_formulation}

Let a matrix  $U={\rm diag}\{\l_1, \dots, \l_n\}$ with distinct $\{\l_j\}$ and the matrices $S, \;\mu,\; R,\; C_\io^r$  related by $S^\iT S^{-1} = (C_\io^r)^{-1} {\rm e}^{2 \pi \i \mu} {\rm e}^{2\pi \i R}C^r_\io$ be given, where 
$\mu$ is diagonal: $\mu={\rm diag}\{\mu_1,\dots,\mu_n\}$; the entries of the matrix $R$ satisfy: $R_{ij} \neq 0$ only if $\mu_i-\mu_j=k$ for a positive integer $k.$

Let $l$ be any oriented admissible line in the $z$-plane passing through the origin ($l=l_+\bigcup l_-$ as before). The complement of the line in the complex plane consists of the left $\Pi^l$ and right $\Pi^r$ half-planes  according to the orientation of the line $l$. 

The Riemann-Hilbert problem we consider in this paper is the problem of finding a matrix function $\Psi(z)$ made up of the following  two functions each defined in one of the two half-planes
\beqn
\label{piecewise}
\Psi(z) = \left\{ \begin{array}{l} \Psi^r(z) , \;\; z \in  \Pi^r \\ \Psi^l(z) , \;\; z \in \Pi^l \end{array}\right.
\eeqn
 such that: 
\begin{itemize}

 \item   ${\rm det} \, \Psi(z)\neq 0$ for $z\in{\mathbb C}\setminus 0.$
 
\item $\Psi$ has an essential singularity at the point at infinity with the asymptotics $\Psi(z) \simeq \left( 1  + {\cal O} ({1/}{z} ) \right) {\rm e}^{zU}$ as $|z|\to\infty$ in $\Pi^r$ or in $\Pi^l;$ $\Psi$ is holomorphic elsewhere in the half-planes. 

\item In a half-disc centered at $z=0$ and contained in $\Pi^r$ the function $\Psi$ has the asym\-ptotics $\Psi^r(z) \simeq  G(z)  z^\mu z^R C_\io^r$ with a  matrix function $G(z)$ holomorphic at the origin. 

\item On  the ray $l_+$  the boundary values of the matrix functions $\Psi^r$ and $\Psi^l$ are related by multiplication with the matrix $S$ from the right, i.e., $\Psi^l(z) = \Psi^r(z)S$ for $z\in l_+.$

\item On the ray $l_-$  the boundary values of the solution are related by multiplication from the right with the matrix $S^\iT,$ i.e., $\Psi^l(z) = \Psi^r(z)S^\iT $ for  $z\in l_- .$

\end{itemize}

We solve this Riemann-Hilbert problem for the matrix $\mu$  being a diagonalization of the matrix $V$ defined by (\ref{V}) -- the entries of $\mu$ are given in Proposition \ref{prop_spectrum} below; the Stokes matrix $S$ is described below in Theorem \ref{thm_Stokes} with examples given in Section \ref{sect_examples}. 
 
\begin{remark}
 {\rm Given a solution $\Psi(z)$ to the Riemann-Hilbert problem formulated in this section,  the matrix function $\hat{\Psi}(z) = \Psi(z){\rm e}^{-zU}$ is holomorphic in the half-planes and solves the  Riemann-Hilbert factorization problem for the function $\hat{S}(z)$ defined on the separating line $l$ by $\hat{S}(z):={\rm e}^{zU}S{\rm e}^{-zU}$ if $z\in l_+$ and $\hat{S}(z):={\rm e}^{zU}S^\iT{\rm e}^{-zU}$ if $z\in l_-.$
   }
\end{remark}

\section{Hurwitz spaces}
\label{sect_Hurwitz} 

\subsection{Definition of Hurwitz spaces}
\label{sect_defHurwitz}

The Hurwitz space $\M_g$ is the set of equivalence classes of ramified coverings  $\lambda:\surf \to \cp,$ where $\surf$ is a compact Riemann surface of genus $g$ and the covering map $\lambda$ is a meromorphic function on $\surf.$ Two coverings  $\lambda:\surf \to \cp$ and $\tilde{\lambda}:\tilde{\surf} \to \cp$ are  equivalent if there exists a biholomorphic map $f : \surf\to\tilde{\surf}$ such that $\tilde{\lambda}{ \circ} f = \lambda.$

We denote the ramification points  by $P_i.$ At these points $d\lambda(P_i)=0.$ Their images $\l_i:=\l(P_i)$ in $\cp,$
the projections of ramification points on the base of the covering, are called the {\it branch points}. A branch point $\l_i$ is called {\it simple} if the differential $d\l$ has a zero of multiplicity one at $P_i$ (i.e., the corresponding ramification point belongs to exactly two sheets of the covering). 
 
The following subspace ${\M}_{g;n_\io,\dots,n_m}$ of the space $\M_g$ is also called the Hurwitz space: ${\M}_{g;n_0,\dots,n_m}$ is the set of equivalence classes of $N$-fold genus $g$ coverings of $\cp$ with simple distinct finite branch points and with the ramification type over the point at infinity  fixed by the numbers $n_\io,\dots,n_m:$
the point at infinity on the base of the covering has $m+1$ preimages, $\lambda^{-1}(\infty) = \{\infty_\io,\dots,\infty_m\};$ the point $\infty_i$ belongs to exactly $n_i+1$ sheets of the covering.

The number $n$ of simple finite branch points is given by the Riemann-Hurwitz formula:  $n=2g+2m+\sum_{i=0}^m n_i.$

Locally in a neighbourhood of a covering of the described type, the set of branch points $\{\l_1,\dots,\l_n\}$ gives coordinates on the Hurwitz space ${\M}_{g;n_0,\dots,n_m}.$ 

The complex structure on the surface $\surf$ is defined by the covering as follows: near a simple ramification point $P_k$ the local parameter is  $x_k(P)=\sqrt{\l(P)-\l_k};$ in a neighbourhood of the point $\infty_i$ with the ramification index $n_i$ the local parameter is given by $ \l(P)^{-1/(n_i+1)}.$

We shall fix a ca\-no\-ni\-cal homology basis $\{a_k;b_k\}_{k=1}^g$ on the surface $\surf,$ i.e.,  we shall work locally in the covering space  $\widehat{\M}_{g;n_0,\dots,n_m}$ whose elements are pairs: a point of the space ${\M}_{g;n_0,\dots,n_m}$ and a canonical basis of cycles on the underlying surface (i.e., a weakly marked Riemann surface and a meromorphic function on it).

\subsection{Bidifferentials on Riemann surfaces}
\label{sect_bidiff}

The bidifferentials described in this section play a key role in the construction of the Frobenius structures on the Hurwitz spaces as will be shown below in Section \ref{sect_HF}. The solution to the system (\ref{RHz}), (\ref{RHi}) will be given in terms of these bidifferentials. 

\paragraph {The bidifferential W.} The {\it canonical meromorphic bidifferential} $W(P,Q)$  is  
defined as the second derivative of the logarithm of the prime form $E(P,Q)$ (see \cite{Fay92}) on a compact Riemann surface: 
\beqn
\label{W-def}
W(P,Q) := d_\iP d_\iQ \log E(P,Q)
\eeqn
with $P$ and $Q$ being points on the surface. $W(P,Q)$ can be alternatively defined as a bidifferential with the following properties: i) it is symmetric; ii) has a second-order pole on the diagonal $P=Q$ with biresidue $1;$ iii) its $a$-periods with respect to either of the arguments vanish:
\beqn
\oint_{a_k}W(P,Q) = 0 , \qquad k = 1,\dots,g .
\label{W-aperiods}
\eeqn
The $b$-periods of $W$ generate the holomorphic differentials $\omega_k$ normalized by $\oint_{a_k}\omega_j=\delta_{jk}:$
$$\!\!\oint_{b_k}\!\!\!W\!(P,Q) = 2 \pi \i \; \omega_k(P) , \hsp k = 1,\dots,g.$$

\paragraph {The deformation of W.}
For a surface $\surf$ of genus $g \geq 1,$  denote by  $\B$ the Riemann matrix ($\B_{ij} = \oint_{b_i}\omega_j$) and  
let $\q$ be a symmetric matrix independent of the branch points $\{ \l_j \}$ and such that the inverse $(\B+\q)^{-1}$ exists. Using such a matrix one can deform the bidifferential $W(P,Q)$ keeping the symmetry property and the singularity structure unchanged. Namely, we introduce the following bidifferential $W_\q(P,Q)$ where $\q$ plays the role of parameter of deformation:
\beqn
\Wq(P,Q) := W(P,Q)-2 \pi \i \sum_{k,l=1}^g(\B+\q)_{kl}^{-1}\omega_k(P)\omega_l(Q).
\label{Wq-def}
\eeqn
In the limit when all diagonal entries of the matrix $\q$ tend to infinity while the off-diagonal entries remain finite, the matrix $(\B+\q)^{-1}$ vanishes and, therefore, the bi\-dif\-fe\-ren\-tial $\Wq$ tends to $W.$  
The normalization condition, analogous to (\ref{W-aperiods}) for $W$,  for the deformed bidifferential has the form:
\beqs
\oint_{b_k} \Wq(P,Q) + \sum_{j=1}^g \q_{jk} \oint_{a_j} \Wq(P,Q) = 0 .
\eeqs
Note that for a fixed symmetric matrix $\q$ the bidiffferential $\Wq$ cannot be defined on an arbitrary Riemann surface: it is not defined for the surfaces satisfying
\beqn
\det\,(\,\B+\q\,) = 0 .
\label{divisor}
\eeqn
This condition may be empty for some matrices $\q,$ for example, it is never satisfied for a real $\q.$ In the general case we do not know how to describe the matrices $\q$ for which the set of surfaces satisfying  (\ref{divisor}) is non-empty. 

\begin{paragraph} {Schiffer and Bergman kernels} are defined for a surface of genus $g\geq 1$ (in genus zero, the Bergman kernel vanishes and the Schiffer kernel coincides with $W$).
The {\it Schiffer kernel} $\Omega(P,Q)$ is a symmetric bidifferential with the singularity of the same type as that of the bidifferential $W.$ It is given by:
\beqn
\Omega(P,Q) := W(P,Q)-\pi\sum_{k,l=1}^g(\imb)_{kl}^{-1}\omega_k(P)\omega_l(Q) .
\label{Omegadef}
\eeqn

The {\it Bergman kernel} $B(P,\Qbar)$  is defined by:
\beqn
B(P,\Qbar):=\pi\sum_{k,l=1}^g(\imb)_{kl}^{-1}\omega_k(P)\overline{\omega_l(Q)}.
\label{Bdef}
\eeqn

The kernels can be alternatively defined as follows \cite{Fay92}. The Schiffer kernel $\Omega(P,Q)$ is a symmetric bidifferential  with a second order pole at the diagonal $P=Q$ with biresidue $1$ and such that the relation
$ {\rm p.v.} \underset{\mathcal{L}}{\iint} \Omega(P,Q) \overline {\omega(P)}
= 0$
 holds for any holomorphic differential $\omega(P)$ on the surface $\surf.$ The Bergman kernel $B(P,\bar{Q})$ is a bidifferential such that 
the integral operator with the kernel   $B(P,\bar{Q})/2\pi i$ acts in the space $L_2^{\scriptscriptstyle{(1,0)}}(\surf)$ of $(1,0)$-forms as an orthogonal projector onto the subspace of holomorphic $(1,0)$-forms.

\subsection{Variational formulas}
\label{sect_Rauch}

Here we study the dependence of the bidifferentials on the point of the Hurwitz space. 
A covering $\lambda:\surf\to\cp$ defines a complex structure on the surface $\surf$ (see Section \ref{sect_Hurwitz}) and this structure depends on the point of the Hurwitz space represented by the coordinates $\{\l_1,\dots,\l_n\}.$ Therefore,
the above bidifferentials defined on  the Riemann surface corresponding to the covering depend on the branch points $\{\l_k\}$ (we consider small variations of $\{\l_k\}$ which keep the canonical homology basis $\{a_k, b_k\}$ on the surface unchanged).

The variational formulas below give the derivatives of the bidifferentials with respect to the simple branch points $\{\l_k\}.$ It is assumed that the projections $\l(P)$ and $\l(Q)$ of the points $P$ and $Q$ on the base of the covering are kept fixed under the differentiation with respect to $\{\l_k\}.$
In the formulas, the fixed quantities are listed after the vertical bar.

For the bidifferential $W,$ the dependence on the simple branch points $\{ \l_k \}$ is given by the Rauch variational formulas \cite{KokKor, Rauch}: 
\beqn
\frac{\d W(P,Q)}{\d\l_k}{\Big\vert}_{\l(P),\l(Q)}=\frac{1}{2}W(P,P_k)W(P_k,Q),
\label{WRauch}
\eeqn
where $W(P,P_k)$ denotes the evaluation of $W(P,Q)$ at  $Q=P_k$ with respect to the local parameter $x_k$:
\begin{equation}
W(P,P_k):=\frac{W(P,Q)}{dx_k(Q)}\Big\vert_{Q=P_k}  \equiv   \left( 2\underset{Q=P_k}{\rm res} \frac{W^2(P,Q)}{d\l(Q)}  \right)^{1/2}.
\label{notation}
\end{equation}
Formulas (\ref{WRauch}) can be easily proved \cite{KokKor} by comparing singularities of the right and left hand sides and using the normalization condition (\ref{W-aperiods}).

Being integrated over $b$-cycles of the surface, the Rauch formulas (\ref{WRauch}) give variational formulas for the holomorphic differentials and the
Riemann matrix:
\beqn
\frac{\d\omega_j(P)}{\d\l_k} {\Big\vert}_{\l(P)}= \frac{1}{2} \omega_j(P_k) W(P,P_k) , \qquad \frac{\d\B_{jl}}{\d\l_k} = \pi  \i \, \omega_j(P_k) \omega_l(P_k) .
\label{Rauch}
\eeqn

Formulas (\ref{WRauch}) and (\ref{Rauch}) allow straightforward computation of the variational formulas for the deformed bidifferential  $\Wq(P,Q)$.  As turns out, they  look formally exactly as those for $W(P,Q):$
\beqn
\frac{\d \Wq(P,Q)}{\d\l_k}{\Big\vert}_{\l(P),\l(Q)}=\frac{1}{2}\Wq(P,P_k)\Wq(P_k,Q) .
\label{WqRauch}
\eeqn
Formulas (\ref{WqRauch}) are only valid at the points of Hurwitz space where the bidifferential is well defined, i.e.,   outside of the divisor (\ref{divisor}) in $\M_{g;n_0,\dots,n_m}.$ 

Note that both bidifferentials $W$ and $\Wq,$ as well as the differentials $\omega_k$ and the matrix $\B,$ are locally holomorphic with respect to branch points $\{ \l_k \},$ i.e., they do not depend on $\{\bar{\l}_k\}$ (see \cite{Fay92}, p. 54). 
However, the Schiffer and Bergman kernels depend also on $\{\lb_k\}$ as can be seen from their explicit definitions (\ref{Omegadef}) and (\ref{Bdef}), where the complex conjugate of the Riemann matrix ${\B}$ enters. 

The variational formulas for $\Omega$ and $B$ can be derived from their definitions    (\ref{Omegadef}) and (\ref{Bdef}) by a straightforward differentiation. They have the form:
\begin{equation}
\begin{split}
\frac{\d\Omega(P,Q)}{\d\l_k}{\Big\vert}_{\l(P),\l(Q)} = \frac{1}{2} \Omega(P,P_k) \Omega(P_k,Q) , 
\qquad
\frac{\d\Omega(P,Q)}{\d\bar{\l}_k}{\Big\vert}_{\l(P),\l(Q)} = \frac{1}{2} B(P,\bar{P}_k) B(Q,\bar{P}_k) , \\
\frac{\d B(P,\bar{Q})}{\d\l_k}{\Big\vert}_{\l(P),\l(Q)} = \frac{1}{2} \Omega(P,P_k) B(P_k,\bar{Q}) ,
\qquad
\frac{\d B(P,\bar{Q})}{\d\bar{\l}_k}{\Big\vert}_{\l(P),\l(Q)} = \frac{1}{2} B(P,\bar{P}_k) \overline{\Omega(P_k,Q)} .
\label{SBRauch}
\end{split}
\end{equation}
The notation here is analogous to that in (\ref{notation}): $\Omega(P,P_k)$ stands for $\!\left( {\Omega(P,Q)}\! / {dx_k(Q)} \right)\! |_{Q=P_k}$ and $B(P,\bar{P}_k) := \left( B(P,\bar{Q}) / \overline{dx_k(Q)} \right) |_{Q=P_k} . $
\end{paragraph}
\vspace{.4cm}

\subsection{Frobenius structures on Hurwitz spaces}
\label{sect_HF}

A Hurwitz space can be endowed with a structure of a Frobenius manifold. In this section we describe the part of the construction of three classes of Frobenius structures on the Hurwitz space $\widehat\M_{g;n_\io, \dots,n_m}$ \cite{2D, doubles, deform} which is needed to introduce the associated Riemann-Hilbert problems. 

\begin{paragraph}{Frobenius manifold.}
A Frobenius manifold \cite {2D} is a complex manifold with a structure of  associative commutative algebra with a unity in the tangent bundle and a {\it Darboux-Egoroff} (flat potential diagonal) metric  compatible with the algebra structure.  
The compatibility means that for a metric $\eta(\cdot,\cdot)$ and any three elements $x,y,w$ of the algebra the condition $\eta(xy,w)=\eta(x,yw)$ holds. (In Frobenius manifold theory, the word ``metric" denotes a bilinear quadratic form.) In addition to this, there are further requirements on the metric. For our purposes it is enough to mention the existence of the {\it Euler vector field} $E$ on the manifold; this field is covariantly linear with respect to the Levi-Civita connection of the metric and acts on the metric by 
\beqn
\label{LieE}
{\rm Lie}_\iE \eta(x,y) = (2-\nu)\eta(x,y),
\eeqn
where $\nu$ is a constant. 

There are two distinguished coordinate sets on the Frobenius manifold: {\it canonical coordinates} denoted by $\{\l_k\}$ and {\it flat coordinates} $\{t_k\}.$ The metric $\eta$ is diagonal in the canonical coordinates:  $\eta = \sum_{i=1}^n g_{ii}(d\l_i)^2,$ where $g_{ii}$ are functions of $\{\l_k\}\,.$ In flat coordinates the coefficients of the metric are constant. 

The multiplication in the  tangent space is diagonal in canonical coordinates: $\d_{\l_i}\d_{\l_j}=\delta_{ij}\d_{\l_i}\,.$ Thus the unit vector field in the algebra is
\beqn
\label{e}
{\bf e}=\sum_{k=1}^n \d_{\l_k}\,.
\eeqn
The Euler vector field in the canonical and flat coordinates has the form \cite{2D}:
\beqn
\label{E}
E := \sum_{k=1}^n \l_k\d_{\l_k} \equiv \sum_{k=1}^n\nu_k t_k \d_{t_k} .
\eeqn
The coefficients $\nu_k$ are called the {\it quasihomogeneity coefficients}.
\end{paragraph}

\begin{paragraph}{A system for rotation coefficients.}
The flat metric on a Frobenius manifold is dia\-go\-nal in the canonical coordinates $\{\l_1,\dots,\l_n\}$ on the manifold. 
For an arbitrary diagonal metric $\eta = \sum_{i=1}^n g_{ii}(d\l_i)^2$ the rotation coefficients are defined by:
\beqn
\label{rot-def}
\beta_{ij} :=\frac{\d_{\l_i}\sqrt{g_{jj}}}{\sqrt{g_{ii}}} , \hsp i\neq j .
\eeqn
The matrices $V$ and $\Gamma$ in equations (\ref{RHz}), (\ref{RHi}) are related by (\ref{V}); they are built from rotation coefficients $\beta_{ij}$ of the flat metric on the Frobenius manifold as follows: $V_{ij}=\beta_{ij}(\l_j-\l_i)$ and $\Gamma_{ij}=\beta_{ij}$ if $i\neq j$ and $V_{ii}=\Gamma_{ii}=0.$

The rotation coefficients $\beta_{ij}$ of a metric on a Frobenius manifold satisfy certain equations as a corollary of the Frobenius manifold axioms. First, they are symmetric:
\beqn
\beta_{ij}=\beta_{ji}.
\label{rotsym}
\eeqn
The symmetry is equivalent to the {\it potentiality} of the metric, i.e., to the existence of a  function $G(\{\l_k\})$ generating the metric coefficients: $g_{ii} = \d_{\l_i}G.$ 
Second, for a Frobenius manifold of dimension $n,$  the rotation coefficients satisfy the following system of differential equations: 
\begin{eqnarray}
&&\d_{\l_k}\beta_{ij} = \beta_{ik} \beta_{jk} , \hsp i,j,k \;\mbox {are distinct}
\label{rot1}\\
&&\sum_{k=1}^n \d_{\l_k} \beta_{ij} = 0
\label{rot2}\\
&&\sum_{k=1}^n \l_k \d_{\l_k} \beta_{ij} = -\beta_{ij} .
\label{rot3}
\end{eqnarray}
The first two equations, by virtue of the Darboux-Egoroff lemma, provide the flatness for the corresponding diagonal potential metric. The third equation (\ref{rot3}) follows from the form  of the action of the Euler field  on the metric (\ref{LieE}).

Equations (\ref{rotsym}) - (\ref{rot3}) provide compatibility condition for the linear system (\ref{RHz}), (\ref{RHi}). 

 \end{paragraph}

\begin{paragraph}{Solution to the system for rotation coefficients on Hurwitz spaces.}
A class of solutions to the system (\ref{rotsym}) - (\ref{rot3}) for rotation coefficients can be constructed on the Hurwitz space in terms of the bidifferentials  introduced in Section \ref{sect_bidiff} as follows.

The Rauch variational formulas (\ref{WRauch}) imply \cite{KokKor} that the following quantities
satisfy equations (\ref{rotsym}) - (\ref{rot3}):
\beqn
\beta_{ij} = \frac{1}{2}W(P_i,P_j) ,\hsp\mbox{for} \hsp i\neq j; \;\;i,j=1,\dots,n .
\label{Wrot}
\eeqn
In particular, the very form of the Rauch formulas (\ref{WRauch}) with $P=P_i$ and $Q=P_j$ coin\-cides with that of equation (\ref{rot1}).
As was first noted in \cite{KokKor}, the rotation coefficients (\ref{Wrot}) correspond to the family of $n$  Frobenius structures of dimension $n$ on Hurwitz spaces found by Dubrovin \cite{2D}.

Due to the similarity of the variational formulas for all bidifferentials from Section \ref{sect_bidiff}, two other sets of rotation coefficients can be found analogously to (\ref{Wrot}) in terms of the other bidifferentials. Namely, the quantities
\beqn
\beta_{ij} = \frac{1}{2}\Wq(P_i,P_j) 
\label{Wqrot}
\eeqn
give rotation coefficients for the so-called deformations \cite{deform} of the Hurwitz Frobenius ma\-ni\-folds from \cite{2D}.

The Schiffer and Bergman kernels define a solution to equations  (\ref{rotsym}) - (\ref{rot3}) on the Hurwitz space considered as a real manifold, i.e., on the space $\M_{g;n_0,\dots,n_m}$ with the local coordinates $\{\l_1,\dots,\l_n;\lb_1,\dots,\lb_n\}.$ Let the indices in equations (\ref{rotsym}) - (\ref{rot3}) run through the set $\{1,\dots,n;\bar{1},\dots,\bar{n}\}$. Then the equations are satisfied by the following quantities: 
\beqn
\beta_{ij} = \frac{1}{2}\Omega(P_i,P_j), \hsp \beta_{i\bar{j}} = \frac{1}{2}B(P_i,\Pbar_j) , \hsp \beta_{\bar{i}\bar{j}} = \frac{1}{2}\overline{\Omega(P_i,P_j)}
\label{SBrot}
\eeqn
with indices $i,j\in \{1,\dots,n\}$ and $\bar{i},\bar{j}\in \{\bar{1},\dots,\bar{n}\}.$ The solution (\ref{SBrot}) defines rotation coefficients for a family of $2n$  Frobenius manifold structures of dimension $2n$ \cite{doubles} on the real Hurwitz space. These manifolds are called the real doubles of Dubrovin's  Frobenius structures on Hurwitz spaces.

\begin{remark}
{\rm Solutions (\ref{SBrot}) are independent of the choice of canonical homology basis on the Riemann surface in contrast to solutions
(\ref{Wrot}) and (\ref{Wqrot}). Therefore, the Frobenius manifolds corresponding to rotation coefficients (\ref{SBrot}) are structures on the Hurwitz space ${\M}_{g;n_0,\dots,n_m}$ whereas the Frobenius structures with rotation coefficients (\ref{Wrot}) and (\ref{Wqrot})  are defined on the covering $\widehat{\M}_{g;n_0,\dots,n_m}$ of the Hurwitz space. }
\end{remark}

\begin{remark}
{\rm  There also exist \cite{deform} deformations of the Schiffer and Bergman kernels. The deformed kernels provide analogously another family of solutions for the system (\ref{rotsym}) - (\ref{rot3}). }
\end{remark}

Here we consider three classes of Riemann-Hilbert problems - those corresponding to the linear systems (\ref{RHz}), (\ref{RHi}) built from the rotation coefficients (\ref{Wrot}), (\ref{Wqrot}) and (\ref{SBrot}).  The solutions to these problems will be given in terms of the bidifferentials $W,$ $\Wq,$ and $\Omega$ and $B,$ respectively. 

\end{paragraph}

\begin{paragraph}{Spectrum of the Frobenius manifolds.}

The monodromy data of the Riemann-Hilbert problem associated to a Frobenius ma\-ni\-fold includes the {\it spectrum} of the ma\-ni\-fold, i.e., the set  $\{\mu_1,\dots,\mu_n\}$ of eigenvalues of the matrix $V$ from (\ref{V}).  The eigenvectors are given by $n$ {\it primary differentials} $\{\phi_j\}$ (the primary differentials are given by integrals of the respective bidifferential - either $W$, or $\Wq,$ or $\Omega$ and $B$ -  with respect to one of the arguments over various contours on the Riemann surface, see \cite{2D, doubles, deform}) evaluated at the ramification points of the covering: 
\beqn
\label{eigenV}
V\vec{\phi}_j = \mu_j \vec{\phi}_j,\hsp \mbox{where} \hsp \vec{\phi}_j = (\phi_j(P_1), \dots, \phi_j(P_n))^\iT .
\eeqn
The evaluation is done with respect to the standard local parameter, similarly to (\ref{notation}). 
A Frobenius manifold is called {\it resonant} if at least one of the differences $\mu_i-\mu_j$ is a nonzero integer. Frobenius manifold structures on Hurwitz spaces are resonant. 
The spectrum is given by the next Proposition.

\begin{proposition}
\label{prop_spectrum}
Frobenius manifolds whose rotation coefficients are given by (\ref{Wrot})  
or by their deformations (\ref{Wqrot}) have the following spectrum $\{\mu_j\}_{j=1}^n:$
\begin{itemize}
\item {$g+m$ values $\mu_j = 1/2$}
\item {$g+m$ values $\mu_j = -1/2$}
\item {$n_i$ values  $\{\mu_j=\frac{\alpha}{n_i+1} - \frac{1}{2}\}_{\alpha=1}^{n_i}$ for every $i = 0, \dots, m$;} 
\end{itemize}
( here $n_i$ is the ramification index at the point $\infty_i$).

The spectrum of the real doubles, the Frobenius manifolds with rotation coefficients (\ref{SBrot}), contains each of the above values twice.
\end{proposition}
{\it Proof.} 
The values $\{\mu_j\}$ are related to the quasihomogeneity coefficients $\{\nu_j\}, \nu$ (see (\ref{LieE}), (\ref{E})) of the Frobenius manifold by \cite{2D}: 
\beqs
\mu_j = 1-\nu_j - \frac{\nu}{2} ;
\eeqs
the coefficients of quasihomogeneity can be found in Proposition 5 of \cite{doubles}.

Alternatively, the spectrum of the Frobenius manifold can be computed by explicitly finding the eigenvalues of the matrix $V$.
In the case of rotation coefficients (\ref{Wrot}) given by $W,$ the primary differentials $\phi_i$ satisfy the following variational formulas \cite{2D, doubles, deform}: 
\beqs
\d_{\l_k}\phi_i(P_j) = \frac{1}{2}W(P_k,P_j)\phi_i(P_k) \hsp \mbox{if} \hsp j\neq k ; \hsp
\mbox{and} \hsp
\sum_{k=1}^n\d_{\l_k} \phi_i(P_j) = 0 .
\eeqs
A short calculation with the help of these formulas  shows that for the vector $\vec{\phi}_i$  from (\ref{eigenV}), the multiplication  by the matrix $V$ is equivalent to the action of the Euler vector field $E$ given by (\ref{E}): $V\vec{\phi}_i = E(\vec{\phi}_i).$ One computes this action by the method used below in the proof of Lemma \ref{Elemma} and thereby finds eigenvalues of the matrix $V$. 

The spectrum of the real doubles can be computed analogously. The quasihomogeneity coefficients and variational formulas for the primary differentials can be found in Proposition 11 of \cite{doubles}.
$\Box$

\end{paragraph}

\section{Solution of the non-Fuchsian Riemann-Hilbert \\ problem }
\label{sect_solution}

In this section we construct a fundamental solution to the system (\ref{RHz}), (\ref{RHi}) associated to a Frobenius manifold structure on a Hurwitz space $\widehat{\M}_{g;n_\io,\dots,n_m}$ and describe the Stokes matrix $S$ for the solution.

For a line dividing the $z$-plane into half planes $\Pi^l$ and $\Pi^r$  we shall find matrix functions $\Psi^l(z)$ and $\Psi^r(z)$ defined in the respective half planes which satisfy equations (\ref{RHz}) and (\ref{RHi}) and have the required asymptotics as $z\to\infty.$ We shall also find  a solution $\Psi_\io(z)$ defined in a disc neighbourhood of $z=0$ with a branch cut such that its monodromy matrix at the origin is in the Jordan canonical form.

The solutions $\Psi^r,$ $\Psi^l$ and $\Psi_\io$ are written in terms of the bidifferential defining the rotation coefficients of the corresponding Frobenius manifold (see Section \ref{sect_HF}). We start with the manifolds with rotation coefficients (\ref{Wrot}) and find the solution in terms of the bidifferential  $W$.

\subsection{A system of contours on a Riemann surface}
\label{sect_contours}

For each $z\in \C\setminus\{0\}$ we consider a linear vector space $\Lambda(z)$ spanned by $n$ differentials of the form 
\beqn
\label{diffspace}
{\rm e}^{z\,\l(Q)}W(Q,P_k)
\eeqn
for $k = 1,\dots,{n} $ defined on the surface.

Here we define a certain space $\Lambda^*(z)$  of equivalence classes of contours on the surface $\surf;$  the integrals of differentials (\ref{diffspace}) over these contours converge.  We shall see later that the pairing given by the integral of the differential form over the contour defines duality between the spaces $\Lambda(z)$ and $\Lambda^*(z).$ 

The space $\Lambda^*(z)$ is the first relative homology of the pair constructed as follows. 
Let us denote by $\tilde{\surf}$ the manifold obtained by blowing up the points $\{\infty_i\}_{i=0}^m$ on the surface $\surf$ into small closed discs ${\cal D}_i.$ 
Let $\hat{\surf}:=\tilde{\surf}\setminus \bigcup_i{\cal D}_i^\io,$ where ${\cal D}_i^\io$ denotes the interior of the disc ${\cal D}_i.$  There is a smooth map $\hat{\surf}\to\surf$ taking the boundary $\d{\cal D}_i\in\hat{\surf}$ to the point $\infty_i\in\surf.$ This map allows us to extend the function $\l(P)$ from the surface $\surf$ to the interior of  $\hat{\surf}.$ Although $\l(P)$ is not defined on the boundary of $\hat{\surf},$ its argument ${\rm arg}\,\l(P)$ is defined also for $P\in\d{\hat{\surf}}.$
Recall now that there are $n_i+1$  sheets of the covering $\l:\surf\to\cp$ glued together at the point $\infty_i.$ Therefore, 
for each $z\in\C\setminus \{0\}$ there are $n_i+1$ arcs in the boundary of the disc ${\cal D}_i$ where $\pi/2<{\rm arg}\{z\l(P) \}<3\pi/2 .$
Let us denote by $\surf^z$ the manifold with a boundary obtained by glueing  the open arcs $\{P\in\d{{\cal D}_i} \mid \pi/2<{\rm arg}\{z\l(P) \}<3\pi/2 \}$ to the interior of $\hat{\surf},$ i.e., $\surf^z:=\hat{\surf}^\io \bigcup_i \{P\in\d{{\cal D}_i} \mid \pi/2<{\rm arg}\{z\l(P) \}<3\pi/2\}.$
Then we define $\Lambda^*(z):=H_1(\surf^z, \d\surf^z).$

Now we shall construct $n$ contours $\{\c_k(z)\}_{k=1}^n$ on $\surf$ along which the exponent in (\ref{diffspace}) is bounded for the given value of $z.$ The equivalence classes corresponding to the contours $\{\c_k(z)\}_{k=1}^n$ in the space $\Lambda^*(z)$ will be shown to form a
 basis in $\Lambda^*(z).$ 

 Let us denote the coordinate on the base of the covering by $\zeta.$
Fix $z\neq 0$ and consider a ray $r_k(z)$ on
the base of the covering  going out of the branch point $\l_k$  in a direction such that for $\zeta \in r_k (z)$
\beqn
 \hsp \hsp \frac{\pi}{2} < {\rm arg} \{ z (\zeta-\l_k) \}   < \frac{3\pi}{2}  
\label{gammaz}
\eeqn
and such that $r_k$ does not pass through any other branch point $\l_j$ with $j\neq k.$
Let us consider the ray $r_k$ as a slit with two banks. Consider an oriented contour $\tilde{\c}_k(z)$ which comes from the point at infinity along one of the banks of the slit $r_k$, makes a small circle around the point $\l_k$ 
and goes back to infinity along the other bank of the slit.

There are two components in the preimage $\lambda^{-1}(\tilde{\c_k}(z))$ which cross a neighbourhood of the ramification point $P_k;$ the sum of these two components is equivalent to a small contour around the point $P_k$, which represents the zero element in the space $\Lambda^*(z).$

We need to specify a choice of one of the two components  in  $\l^{-1}(\tilde{\c_k}(z)).$ This choice  affects the overall sign in the formula (\ref{Psi_rl}) below for the solution with the given asymptotics (\ref{formal}), (\ref{atinf}) in a neighbourhood of the irregular singularity. 

In the local parameter $x_k(P) = \sqrt{\l(P)-\l_k}$ near $P_k$ the requirement (\ref{gammaz}) takes the form: ${\pi}/{2} < {\rm arg} \{z\, x_k^2  \}  < {3\pi}/{2} .$
This condition holds in two sectors in the $x_k$-plane:
\beqn
 \sigma_{\pm}:=\left\{ x_k \mid \frac{\pi}{4} - \frac{{\rm arg}\,z}{2} < {\rm arg} \{ \pm x_k  \}  <   \frac{3\pi}{4} - \frac{{\rm arg}\,z}{2} \right\}.
\label{gammax}
\eeqn

Let us choose the contour $\c_k(z)$ to be (in the equivalence class of) that component of  $\l^{-1}(\tilde{\c_k}(z))$ which approaches $P_k$ in the sector $\sigma_+$ and goes away from $P_k$ in the sector $\sigma_-.$ 

Thus the contour $\c_k(z)$ 
starts at the point $\infty_i$, goes round $P_k$ and ends at $\infty_j$ for some, possibly equal, $i$ and $j.$
A change of the direction in which the contour $\c_k(z)$ winds around the point $P_k$ does not change the integral of a differential (\ref{diffspace}) over the contour due to the vanishing of the residues of the differentials at $P_k.$

The variation of   the ray $r_k(z)$ on the base keeps the corresponding contour $\c_k(z)$ in the same equivalence class as long as the ray $r_k(z)$ remains in the sector (\ref{gammaz}) and stays away from other branch points (i.e., if the deformation of the ray to a new position does not meet any branch points).
Moreover, for finite values of $z$, the contour can be deformed in a neighbourhood of $P_k$ as long as requirement (\ref{gammaz}) is satisfied for projections of the end-parts of the contour lying in a neighbourhood of the point at infinity. 
In what follows we shall speak about contours on the surface $\surf$ meaning the elements of the space $\Lambda^*(z)$ which they represent.

 Note  also that projections $\l(\c_k(z))$ of the contours $\c_k(z)$ on the base of the covering  do not depend on small variations of the branch points $\{\l_j\}$ of the covering.

\subsection{Construction of the solution}
\label{sect_infty}
  
Let $\phi$ be the angle between the admissible line $l$ and the real axis. Let us define the contours $\{\c_k^r\}$ along which the forms (\ref{diffspace}) are bounded for all $z\in\Pi^r.$ The contours $\{\c_k^r\}$ are equivalent to the contours  $\{\c_k(z)\}$ for some $z\in\Pi^r$ and their projections on the base of the covering do not depend on $z.$ Namely, for a point $Q$ on the contour $\c_k^r$ away from a neighbourhood of $P_k$ we require:
\beqs
{\rm arg}(\l(Q)-\l_k) = \frac{3\pi}{2} - \phi .
\eeqs

Analogously, the set of contours $\{\c_k^l\}$ is formed by fixing the direction so that the exponents in (\ref{diffspace}) are bounded along the contours for any value of $z$ in $\Pi^l:$ for $Q \in \c^l_k$ away from a neighbourhood of $P_k$

\beqs
{\rm arg}(\l(Q)-\l_k) = \frac{\pi}{2} - \phi .
\eeqs

Recall from the definition of contours $\c_k$ (see inequality (\ref{gammax}) and the comment after it) that  the contour $\c^r_k$ crosses a neighbourhood of  $P_k$ so that in the local parameter $x_k$ it passes from the sector where ${\rm arg}(x_k)= {3\pi}/{4} - \phi/2$  to the sector where ${\rm arg}(-x_k)= {3\pi}/{4} - \phi/2.$ 
 A similar condition for $\c_k^l$ is induced by the definition of $\c_k(z).$

Thus the integrals of differentials (\ref{diffspace}) over contours $\c^r_k$ (respectively, $\c^l_k$) are defined for any $z$ in the half-plane $\Pi^r$ (respectively, $\Pi^l$). These integrals provide solutions $\Psi^{r/l}$ for the linear system:

\begin{theorem}
\label{thm_Psi_rl}
Let the contours $\c_j^{r}$ and $\c_j^{l}$ be as described above.
The following matrix functions $\Psi^l = (\Psi^l_{ij})$ and $\Psi^r= (\Psi^r_{ij})$ defined in the half-planes $\Pi^l$ and $\Pi^r,$  respectively, satisfy equations (\ref{RHz}) and (\ref{RHi}):
\beqn
\Psi^{l/r}_{ij} (z) :=\frac{1}{2\i\sqrt{\pi}}  \frac{1}{\sqrt{z}} \int_{\c^{l/r}_j} {\rm e}^{z\,\l(Q)}W(Q,P_i) .
\label{Psi_rl}
\eeqn
These solutions have the following asymptotics as $z$ tends to infinity:
\beqn
\Psi^{l/r}(z) = \left( 1 + {\cal O} ( {1}/{z} ) \right) {\rm e}^{zU} , \hsp z \to \infty, \;\; z \in \Pi^{l/r} ,
\label{asymptotics}
\eeqn
where $U$ is the diagonal matrix $U=diag(\l_1,\dots,\l_n) .$
\end{theorem}

Before proving the theorem let us  look at the action of  the unit vector field (\ref{e}) and the Euler vector field (\ref{E}) on the integrals from (\ref{Psi_rl}) 
formulated in the following two lemmas. 
\begin{lemma}
\label{elemma}
Consider a covering of $\cp$  with simple finite branch points $\{\l_k\}$.
 Let $\c_j$ be one of the contours $\c_j^r$ or $\c_j^l.$  Then the following relation holds: 
\beqn	
\sum_{k=1}^n \d_{\l_k} \int_{\c_j} {\rm e}^{z\,\l(Q)}W(Q,P_i) = z \int_{\c_j} {\rm e}^{z \, \l(Q)}W(Q,P_i) .
\label{ePsi0}
\eeqn
\end{lemma}
{\it Proof.}
To compute the sum of partial derivatives in the left hand side we use the identity $\sum_{k=1}^n \d_{\l_k} h(\{\lambda_i\})= \frac{d}{d\delta}|_{\delta=0} h(\{\lambda_i + \delta \})$, where $h$ is a function of the branch points. 

Consider a biholomorphic map of the Riemann surfaces $\surf \to \surf^\delta$ which acts in every sheet of  $\surf$ by sending the point $P$ with the projection $\l(P)$  to the point $P^\delta$  projecting to $\l(P^\delta)=\l(P)+\delta$ on the base. The branch points $\{\l_i\}$ are then mapped to $\{\l_i+\delta\}.$ 

The bidifferential $W$ stays invariant under biholomorphic mappings of the surfaces, therefore the equality $W(P,Q) = W^\delta(P^\delta,Q^\delta)$ holds, where $W^\delta$ is the bidifferential $W$ defined on $\surf^\delta.$ Since the local parameters $x_i(P) = \sqrt{\l(P)-\l_i}$ are invariant under the mapping, this equality also holds if one of the arguments of the bidifferential coincides with a branch point (see (\ref{notation}) for definition of the differential $W(P,P_i)$):
\beqn
\label{Wdelta}
W(P,P_i) = W^\delta(P^\delta,P_i^\delta) .
\eeqn

For the quantity $\int_{\c_j} {\rm e}^{z\,\l(Q)}W(Q,P_i)$ we have 
\beqs
\sum_{k=1}^n\d_{\l_k} \int_{\c_j} {\rm e}^{z\,\l(Q)}W(Q,P_i) = \frac{d}{d\delta}\big{\vert}_{\delta=0}\int_{\c_j} {\rm e}^{z\,\l(Q)}W^\delta(Q,P^\delta_i) ,
\eeqs
which after changing the variable of integration and then using (\ref{Wdelta}) proves the lemma:
\beqs
 = \frac{d}{d\delta}\big{\vert}_{\delta=0}\int_{\c_j} {\rm e}^{z\,(\l(Q)+\delta)}W^\delta(Q^\delta,P^\delta_i) = \frac{d}{d\delta}\big{\vert}_{\delta=0}\int_{\c_j} {\rm e}^{z\,(\l(Q)+\delta)}W(Q,P_i) =  \! z \! \int_{\c_j} {\rm e}^{z\,\l(Q)}W(Q,P_i) .
\eeqs
Here we used the invariance of the contours $\c_j$ under the mapping $\surf\to\surf^\delta$.
$\Box$

\begin{lemma}
\label{Elemma}
In the settings of Lemma \ref{elemma}, the action of the Euler vector field (\ref{E}) on the integrals is given by 
\beqn
\label{EPsi0}
\sum_{k=1}^n \l_k \d_{\l_k} \int_{\c_j} {\rm e}^{z\,\l(Q)}W(Q,P_i)= z\int_{\c_j} \l(Q){\rm e}^{z\,\l(Q)}W(Q,P_i)-\frac{1}{2}\int_{\c_j} {\rm e}^{z\,\l(Q)}W(Q,P_i) .
\eeqn
\end{lemma}
{\it Proof.}
Consider a biholomorphic map from the surface $\surf$ to the surface $\surf^\epsilon$ which acts in each sheet of the covering by $P\mapsto P^\epsilon$ where the point $P^\epsilon$ is such that $\l(P^\epsilon)= (1+\epsilon)\l(P).$ The bidifferential $W$ is invariant under biholomorphic maps, hence $W^\epsilon(P^\epsilon,Q^\epsilon)=W(P,Q);$ the local parameter $x_k(P)=\sqrt{\l(P)-\l_k}$ becomes $x^\epsilon_k (P^\epsilon)= x_k(P)\sqrt{1+\epsilon}.$ Therefore $W(Q,P_i)$ transforms as follows
\beqn
\label{Wepsilon}
W^\epsilon(Q^\epsilon,P^\epsilon_i):= \frac{W^\epsilon(Q^\epsilon,P^\epsilon)}{dx^\epsilon_i(P^\epsilon)}\big{\vert}_{P^\epsilon=P^\epsilon_i} = \frac{1}{\sqrt{1+\epsilon}}W(Q,P_i) .
\eeqn

For a function of branch points we have $(\sum_k\l_k\d_{\l_k}) h(\{\l_i\}) = \d_\epsilon|_{\epsilon=0}h^\epsilon(\{(1+\epsilon)\l_i\}).$ Thus the left hand side of (\ref{EPsi0})  becomes:
\beqs
\sum_{k=1}^n \l_k \d_{\l_k} \int_{\c_j} {\rm e}^{z\,\l(Q)}W(Q,P_i) = \frac{d}{d\epsilon}\big{\vert}_{\epsilon=0} \int_{\c_j} {\rm e}^{z\,\l(Q)} W^\epsilon(Q,P^\epsilon_i) . 
\eeqs
The contours of integration do not change under the map $\surf \to \surf^\epsilon$. Therefore changing the variable of integration $Q$ to $Q^\epsilon$ and using (\ref{Wepsilon}), we prove the lemma: 
\begin{multline*}
\sum_{k=1}^n \l_k \d_{\l_k} \int_{\c_j} {\rm e}^{z\,\l(Q)}W(Q,P_i) = \frac{d}{d\epsilon}\big{\vert}_{\epsilon=0} \int_{\c_j} {\rm e}^{z\,\l(Q^\epsilon)} W^\epsilon(Q^\epsilon,P_i^\epsilon)  \\
= \frac{d}{d\epsilon}\big{\vert}_{\epsilon=0} \int_{\c_j} {\rm e}^{z\,(1+\epsilon)\l(Q)}\frac{W(Q,P_i)}{\sqrt{1+\epsilon}}= z\int_{\c_j} \l(Q){\rm e}^{z\,\l(Q)}W(Q,P_i)-\frac{1}{2}\int_{\c_j} {\rm e}^{z\,\l(Q)}W(Q,P_i) .
\end{multline*}
$\Box$

{\it Proof of Theorem \ref{thm_Psi_rl}.}
To prove the first part of the theorem, we shall verify that each column  of the matrices $\Psi^r$ and $\Psi^l$ satisfies equation (\ref{RHz}). 
The $(ij)$-entry of the second term in the right-hand side of (\ref{RHz}) for $\Psi = \Psi^{r/l}$ has the form: 
\beqs
\frac{1}{z}\left(V\Psi\right)_{ij} =\frac{1}{2\i\sqrt{\pi}} \frac{1}{\sqrt{z^3}}   \sum_{k=1, k\neq i}^n \frac{1}{2} W(P_i,P_k)(\l_k-\l_i) \int_{\c_j} {\rm e}^{z\,\l(Q)}W(Q,P_k) ,
\eeqs
which, by virtue of the Rauch variational formulas (\ref{WRauch}) for $W,$ rewrites as 
\beqs
\frac{1}{z}\left(V\Psi\right)_{ij} = \frac{1}{2\i\sqrt{\pi}}\frac{1}{\sqrt{z^3}} \left[  
\sum_{k=1, k\neq i}^n \!\! \lambda_k\d_{\lambda_k} \int_{\c_j} {\rm e}^{z\,\l(Q)}W(Q,P_i) -\lambda_i \sum_{k=1, k\neq i}^n \!\! \d_{\lambda_k} \int_{\c_j} {\rm e}^{z\,\l(Q)}W(Q,P_i)  \right] .
\eeqs
Expressing the sum of derivatives in the second term with the help of Lemma \ref{elemma}  yields
\beqs
\frac{1}{z}\left(V\Psi\right)_{ij} = \frac{1}{2\i\sqrt{\pi}}\frac{1}{\sqrt{z^3}} \left[E \int_{\c_j}  {\rm e}^{z\,\l(Q)}W(Q,P_i)  -\l_i z \int_{\c_j}  {\rm e}^{z\,\l(Q)}W(Q,P_i) \right] ,
\eeqs
where $E$  is the Euler vector field (\ref{E}). Using now Lemma \ref{Elemma} for the action of the Euler vector field, we obtain:
\beqs
\frac{1}{z}\left(V\Psi\right)_{ij} = \frac{1}{2\i\sqrt{\pi}}\frac{1}{\sqrt{z}} \left[\!\int_{\c_j}\! \l(Q){\rm e}^{z\,\l(Q)}W(Q,P_i)\!-\!\frac{1}{2z}\int_{\c_j} \! {\rm e}^{z\,\l(Q)}W(Q,P_i) \! -\! \l_i  \int_{\c_j} \! {\rm e}^{z\,\l(Q)}W(Q,P_i) \! \right] .
\eeqs
Thus the $(ij)$-entry of the matrix in the right-hand side of (\ref{RHz}) with $\Psi = \Psi^{r/l}$ is  given by 
\beqs
\left(U\Psi + \frac{1}{z}V\Psi\right)_{ij} =\frac{1}{2\i\sqrt{\pi}} \frac{1}{\sqrt{z}} \left[\int_{\c_j} \l(Q){\rm e}^{z\,\l(Q)}W(Q,P_i)-\frac{1}{2z}\int_{\c_j} {\rm e}^{z\,\l(Q)}W(Q,P_i)  \right] ,
\eeqs
which coincides with $(\d_z\Psi^{r/l})_{ij}.$ This proves that the matrices $\Psi^{r/l}$ (\ref{Psi_rl}) satisfy equation (\ref{RHz}). 

The fact that the matrices (\ref{Psi_rl}) satisfy equation (\ref{RHi}) follows from the compatibility of the system (\ref{RHz}), (\ref{RHi}). One can also verify this directly by using Lemma \ref{elemma} and the Rauch variational formulas (\ref{WRauch}) for the bidifferential $W.$

The asymptotics (\ref{asymptotics}) can be computed with the help of the saddle-point integration method (see \cite{Evgrafov, Wasow}) as follows.

Let us consider the matrix function $\Psi^{l}(z)  {\rm e}^{-zU} $ and prove that it behaves as $\Id + {\cal O} ( {1}/{z} )$ in the limit as $z\to\infty$ in $\Pi^l.$ The entries of this matrix have the form: 
\beqn
\label{new}
\left(  \Psi^{l}(z)  {\rm e}^{-zU}  \right)_{ij}=\frac{1}{2\i\sqrt{\pi}}  \frac{1}{\sqrt{z}} \int_{\c^{l}_j} {\rm e}^{z\,(\l(Q)-\l_j)}W(Q,P_i) .
\eeqn

The saddle-point integration method is based on the idea that the asymptotics of the integral (\ref{new}) as $z\to \infty$ estimates by the sum of contributions of the points on the integration contour such that for any fixed $z\in\C\setminus \{ 0 \}:$ i) the complex modulus of the exponent under the integral attains at this point its local maximum over the contour; ii) the maximum cannot be made smaller by any small deformation of the integration path. 

As is shown in \cite{Evgrafov}, Theorem 1.6.1, these requirements are achieved for a point of maximum of the function ${\rm Re}\{z (\l(P)-\l_j)\}$ over the integration contour if and only if it coincides with a saddle point of the function (points where the derivative of $(\l-\l_j)$ with respect to local parameter vanishes are saddle points of the harmonic function ${\rm Re}\{z (\l(P)-\l_j)\}$).

 For the integral (\ref{new}) there is only one such point:  $Q=P_j$. For $i\neq j$ the path of integration $\c^{l}_j$ can be deformed to pass through the ramification point $P_j.$
Therefore, in the case of the integrals (\ref{new}) with distinct $i$ and  $j$  one can apply the following asymptotic estimates  obtained by the saddle-point method, see \cite{Evgrafov} paragraph $1.6$. 
The contribution of the point $P_j$ to the integral in (\ref{new}) for $i\neq j$ is given by: $\sqrt{-{\pi}/{z}} \left( W(P_j,P_i) + {\cal O}(1/z) \right)$. Thus, for the off-diagonal entries of the matrix (\ref{new}) we have $\left(  \Psi^{l}(z)  {\rm e}^{-zU}  \right)_{ij} = {\cal O} (1/z)$ as $z\to \infty.$

In the expression for diagonal terms of the matrix (\ref{new}), the integrand is singular at the saddle point of the function  ${\rm Re} \{\!z(\l - \l_j)\!\}$ so the integration contour cannot be deformed to pass through it. However, the asymptotics of the integral as $z\to \infty$ is also determined by the integral over the part of the contour lying in a small disc $D_j$ centered at $P_j:$ the integral over the remaining part of the contour is exponentially small. In the disc $D_j$ the integral from (\ref{new}) has the form: 
\beqn
\label{tempexp}
\int_c   {\rm e}^{\,z\,x_j^2}\left( \frac{1}{x_j^2} + {{\cal O}(1)} \right) dx_j , 
\eeqn 
where $z\in \Pi^l$ and $c$ is an arc in the $x_j$-plane starting on the ray ${\rm arg} \, x_j =  \pi/4-\phi/2 $ ending on the opposite ray ${\rm arg}\, x_j =5 \pi/4-\phi/2$ and not passing through the origin. For simplicity, we take the start and end point of $c$ to be ends of a diameter of $D_j$ and denote them by $\pm r$ with ${\rm Re} \,r>0.$ The contribution of the holomorphic part of the expansion of $W(P,P_j)$ in (\ref{tempexp})  to the integral vanishes in the limit $z\to \infty.$  The integral $\int_c   {\rm e}^{\,z\,x_j^2}/{x_j^2}  \,dx_j $ after a change of variables $u^2:=-zx_j^2$ and integration by parts yields:
\beqn
\label{tempint}
 -2\sqrt{-z} \int_{ c\sqrt{-z}} {\rm e}^{-u^2}du +\frac{2}{r}{\rm e}^{z r^2} .
\eeqn
Since ${\rm Re}\{z r^2\}$ is negative for $z\in \Pi^l,$ the second term vanishes in the limit $z\to \infty.$ The  first term reduces to the Gaussian integral as follows.  Let us consider two vertical segments in the $u$-plane going from the ends of the contour $c\sqrt{-z}$ to the real line. Note that the segments lie in the domain of the $u$-plane where ${\rm Re}\{- u^2\}$ is negative for $z\in\Pi^l.$
On one of the segments we have $u= r\sqrt{-z} \pm\i y$ with $0<y<|{\rm Im}\{r\sqrt{-z}\}|.$  The modulus of the integral over this segment can be  estimated from above by the quantity $|r\sqrt{-z}|{\rm e}^{{\rm Re}\{zr^2\}},$ which tends to zero as $z\to\infty.$ The integral of ${\rm e}^{-u^2}du$ over the second segment is estimated similarly. Therefore, we conclude that the integral in (\ref{tempint}) in the limit $z\to\infty$ coincides with the  Gaussian integral $-\int_\R {\rm e}^{-u^2}du = -\sqrt{\pi}.$ 
Thus,  the quantity in (\ref{tempint}) tends to $2\sqrt{-z}\sqrt\pi$ as $z\to\infty$ in $\Pi^l.$ Therefore, for the diagonal entries of the matrix (\ref{new}) we have $\underset{z\to\infty}{\rm lim} \left(  \Psi^{l}(z)  {\rm e}^{-zU}  \right)_{ii} =1.$
$\Box$

\begin{theorem}
\label{thm_det}
The determinant of the solution (\ref{Psi_rl}) to system (\ref{RHz}), (\ref{RHi}) is given by:
\beqn
\label{det}
{\rm det} \; \Psi^{l/r} = {\rm exp}\{z\sum_{k=1}^n \lambda_k\} .
\eeqn
\end{theorem}
{\it Proof.}  The formula ${\rm tr} \{\left(\partial_z A\right) A^{-1} \} =\left(\partial_z  {\rm det A}\right)  ({\rm det} A)^{-1}$ holds for any matrix function $A(z).$ Applying it  to $\Psi(z)$, from the form of equations (\ref{RHz}), (\ref{RHi}) one obtains (\ref{det}) up to a constant factor. The asymptotics (\ref{asymptotics}) of the solution implies that the factor is equal to one. 
$\Box$

\begin{remark}
\label{rmk_basis}
{\rm Theorems \ref{thm_Psi_rl} and \ref{thm_det} imply that the contours $\{\c_k^r\}$ (or, analogously, $\{\c_k^l\}$ ) form a basis in the space $\Lambda^*(z)$ dual to the space of differentials (\ref{diffspace}), where $z$ is in the right (left) half-plane.  
}
\end{remark}

\begin{remark} {\rm Let us briefly discuss the relationship of our formula (\ref{Psi_rl}) to previous results of Dubrovin and Krichever.
The contours on the Riemann sphere given by the projections $\l(\c_k^r)$ and $\l(\c_k^l)$ were used in \cite{DubrovinPainleve}, Lecture 5 for describing the fundamental solution to (\ref{RHz}) having the asymptotics (\ref{asymptotics}) at infinity in terms of solutions to an auxiliary Fuchsian system of ODE (whereas in our case the form we integrate is defined by purely algebraic data).  Moreover,  Dubrovin gives formulas for flat coordinates of the deformed connection on the Frobenius manifold in \cite{DubrovinPainleve}, formulas (5.62) and (5.69), and the relationship of these coordinates to the function $\Psi$ in \cite{2D}, formula (3.119). Using these relations and the formalism of the author's paper \cite{doubles} which expresses all ingredients of Dubrovin's construction in terms of the bidifferential $W$ (\ref{W-def}),  it is possible to arrive to formulas (\ref{Psi_rl}) in an alternative way. 
 
In Krichever's paper \cite{Krichever} solutions to the WDVV equations which generically do not satisfy 
the quasihomogeniety condition were studied. In Theorem 4.5 of \cite{Krichever} a solution to a linear system which is equivalent  to our isomonodromy conditions (\ref{RHi}) rewritten in the flat coordinates on the Frobenius manifold was  given (without proof). However, Krichever's  solution is only formal: no integration contours are given nor is the completeness of the set of the solutions  discussed.
}
\end{remark}

\begin{proposition}
\label{prop_cont}
The solution $\Psi^{r}$ (respectively $\Psi^{l}$) given by Theorem \ref{thm_Psi_rl}  admits analytical continuation preserving the asymptotics (\ref{asymptotics}) into the smallest sectorial neighbourhood of the right (respectively left) half-plane bounded by a pair of the Stokes rays. 
\end{proposition}

{\it Proof.}
The projection of the contour $\c_k^r$ in (\ref{Psi_rl}) is fixed in such a way that the quantity ${\rm Re} \{z(\l(P)-\l_k)\}$ is negative for all values of $z$ in the right half-plane $\Pi^r.$ The validity of this condition implies, by virtue of Theorem \ref{thm_Psi_rl}, the required asymptotic behaviour (\ref{asymptotics}) for the solution $\Psi^r.$ If the values of $z$ are restricted to a small sector in $\Pi^r$ adjacent to the ray $l_+$ then the integration contour $\c_k^r$ can be deformed to include the sector of the left half-plane between $l_+$ and the next Stokes ray into the condition ${\rm Re} \{z(\l(P)-\l_k)\}<0.$  Namely, consider the sector containing the ray $l_+$:
\beqn
\label{sectneigh}
 \phi -\varepsilon < {\rm arg } \,z < \phi+\varepsilon ,
\eeqn
where $\varepsilon>0$ and $\phi$ is the angle between the ray $l_+$ and the real line. 
Let the contour $\hat{\c}_k^r$ be obtained from $\c_k^r$ by a clockwise $\varepsilon$-turn about the point $P_k:$ 
for $P$ belonging to $\hat{\c}^r_k,$  the equality 
\beqn
\label{cktilde}
{\rm arg}(\l(P)-\l_k) = \frac{3\pi}{2} - \phi -\varepsilon 
\eeqn
holds.
Then the matrix function $\Psi^r_\varepsilon$ given by the expression (\ref{Psi_rl}) with the integration contours replaced by the new contours $\{\hat{\c_k^r}\}$ gives the analytic continuation of the solution $\Psi^r$ into the sector $ \phi < {\rm arg } \,z < \phi+\varepsilon$ in the left half-plane. The new solution has the required asymptotics (\ref{asymptotics}) as $|z| \to \infty$ in the sector (\ref{sectneigh}) as follows from the proof of Theorem \ref{thm_Psi_rl}. As is easy to see, in the part of the sector (\ref{sectneigh}) lying in the right half-plane the integrals (\ref{Psi_rl}) over contours $\c_k^r$ and $\hat{\c_k^r}$ coincide: $\Psi^r_\varepsilon(z) = \Psi^r(z)$ for $z$ such that $ \phi -\varepsilon < {\rm arg } \,z < \phi.$

The parameter $\varepsilon$ can be increased till the ray ${\rm arg } \,z = \phi+\varepsilon$ meets a Stokes ray $r_{ij}$. 
After that the analytical continuation preserving the asymptotics (\ref{asymptotics}) breaks down since the set of contours $\{\hat{\c_k^r}\}$ (\ref{cktilde}) is no longer a smooth deformation of the set $\{\c_k^r\}:$ 
there is $k$ such that the contour $\hat{\c}_k^r$ can no longer be obtained by turning the contour $\c_k^r$ about $P_k$ since a branch cut joining the points $P_i$ and $P_j$ will be in the way of such a deformation. 

The analytic continuation of $\Psi^r$ beyond the ray $l_-$ and the analytic continuation of $\Psi^l$ are made analogously. 
$\Box$

From the form of solutions  (\ref{Psi_rl}) one sees that a transformation which takes one of the matrix functions $ \Psi^{l},\; \Psi^{r}$ to another one amounts to the transformation between the respective systems of integration contours on the Riemann surface.  

Namely, the analytical continuation of the solutions $\Psi^r$ and $\Psi^l$ (\ref{Psi_rl}) into a small sector containing the ray  $l_+$ inside is made by the deformation of the integration contours $\c_k^r$ and $\c_k^l$ into the contours $\hat{\c}_k^r$ and $\hat{\c}_k^l$ as described in the proof of Proposition \ref{prop_cont}. The contours $\hat{\c}_k^r$ and $\hat{\c}_k^l$ belong to the space $\Lambda^*(z_+)$ for  $z_+\in l_+.$ The two system of contours $\{\hat{\c}_k^r\}_{k=1}^n$ and $\{\hat{\c}_k^l\}_{k=1}^n$ give two bases in the space $\Lambda^*(z_+),$ therefore they are related by a linear transformation.  This transformation is given by the Stokes matrix (\ref{SM}) of equation (\ref{RHz}).

\renewcommand{\thefootnote}{\arabic{footnote}}

In other words, the following theorem holds. 
\begin{theorem}\footnote{ This theorem and Theorem \ref{thm_monodromy} below were conjectured by C. Hertling (private communication).}
\label{thm_Stokes}
Consider a  Frobenius manifold structure on the Hurwitz space and the corresponding linear system (\ref{RHz}). Let $l = l_-\bigcup l_+$ be an admissible line in the $z$-plane for the system.
Consider two systems of contours $\{\hat{\c}^r_j\}$ and $\{\hat{\c}^l_j\}$ on the covering of the Riemann sphere defined above, which form two bases in the space $\Lambda^*(z_+)$ with $z_+\in l_+.$  Then  the matrix $S$ of transformation between the two bases
\begin{equation*}
(\hat{\c}_1^l, \dots, \hat{\c}_n^l) = (\hat{\c}_1^r, \dots, \hat{\c}_n^r)S
\end{equation*}
is the Stokes matrix of the Frobenius manifold. 
\end{theorem}

Examples of computation of the Stokes  matrix will be given in Section \ref{sect_examples}.

\subsection{Monodromy matrix at zero: Jordan canonical form}
\label{sect_origin}

Alternatively to dividing a neighbourhood of $z=0$ into two domains $\Pi^r$ and $\Pi^l,$ one can consider a disc neighbourhood of the origin slit along the segment of the ray $l_-$ lying in the disc - we denote this domain by $D.$

In this section we shall construct a fundamental solution $\Psi_\io$  defined in $D$ whose mo\-nod\-romy $M_{\io}$ around the origin is given by a matrix in the Jordan canonical form. The asymptotics of the solution $\Psi_\io$ at $z=0,$ similarly to the asymptotics (\ref{Psi_r_at0}) of $\Psi^r,$ has the form:
\beqs
\Psi_\io(z) \simeq  G(z)  z^\mu z^R C_\io , \qquad z\sim 0,
\eeqs
where $C_\io$ is a constant matrix. 
On the overlap of the domain $D$ and the right half-plane $\Pi^r$ the solutions $\Psi_\io$ and $\Psi^r$ are related by  
\beqn
\label{CM}
\Psi_\io(z) = \Psi^r(z)C,
\eeqn
 where $C$ is called the {\it connection matrix.} Thus the constant matrices in the asymptotics of the solutions $\Psi^r$ and $\Psi_\io$ at the origin  satisfy $C_\io=C_\io^r C.$ The matrices $M_{\io}$ and $C$ are related to the Stokes matrix and to the monodromy $M$ of the solution $\Psi^r$ around the origin by $M_{\io}=C^{-1}S^\iT S^{-1}C\equiv C^{-1}M C.$

We start with constructing an auxiliary solution $\tilde{\Psi}_\io$  given by the integrals of the form (\ref{Psi_rl}) over a natural system of contours on the Riemann surface. These contours denoted by $\{\gamma_k(z)\}_{k=1}^n$ are defined as follows: 

\begin{itemize}

\item {for $k=1, \dots, 2g$ the contours $\gamma_k$ are given by  $a$- and $b$-cycles on the Riemann surface (the canonical homology basis of $\surf$).} 

\item for $k = 2g+1, \dots, 2g+m$ the contours $\gamma_k$ are the $m$ cycles encircling points $\infty_1,\dots,\infty_m$ counterclockwise;  
we shall also denote these cycles by  $\{{\calV}_i\}_{i=1}^m.$
 
\end{itemize}

Consider a loop on the base of the covering encircling the point $\zeta=\infty$ counterclockwise such that all points $\lambda_k$ lie outside of the loop.
Corresponding to this loop there is a permutation of sheets of the covering, which we denote by $\sigma.$
Assume also that the sheets are ordered so that the point $\infty_\io$ belongs to the sheets from the $0$th to the $n_\io$th; the point $\infty_i$ belongs to the sheets from the $k_i$th to $(k_i+n_i)$th where $k_i:=\sum_{j=0}^{i-1}(n_j+1).$

The contours from the next two groups start and end at the points on the covering projecting to the point at infinity on the base. 
The direction in which the contours  leave and approach these points  depends on ${\rm arg} \{z\}$ and is determined by the condition for 
$Q\in\lef_k(z),$ $k>2g+m:$
\beqn
\label{gammaz1}
\frac{\pi}{2} < {\rm arg} \{ z \l(Q)  \}  < \frac{3\pi}{2} .
\eeqn

\begin{itemize}
 
\item for $k=2g+m+1,\dots,2g+2m$ the contours $\gamma_k$ are given by the $m$ paths (denoted by ${\calW}_{\io i}(z),\, i=1,\dots,m$) connecting $\infty_\io$ with $\infty_i$, $i\neq 0,$ approaching the endpoints in the direction fixed by (\ref{gammaz1}) on the $0$th and $k_i$th sheets, respectively.

For definiteness in the choice of the contours let us connect all points $\infty_i$ by a curve in the fundamental polygon of the surface. Then we require the contours $\calW_{\io i}(z)$ to lie inside the fundamental polygon and not to cross the curve connecting the points $\infty_i .$

\item
for $k=2g+2m+1,\dots,n$ where $n=2g+2m+\sum_{i=0}^m n_i:$ 
for any $i\in\{0,\dots,m\}$ we take $n_i$ contours ${\calT}_{i;\alpha}(z)$ with  $\alpha=1,\dots,n_i .$ The contour $\calT_{i;\alpha}(z)$ leaves $\infty_i$ along a  direction satisfying (\ref{gammaz1})  on the sheet number $\sigma^{\alpha-1}(k_i)$ at $\infty_i$, goes counterclockwise around the point $\infty_i$ to the sheet number $\sigma^\alpha(k_i),$ and there comes back to $\infty_i$ in the same direction  (projection of  ${\calT}_{i;\alpha}(z)$ on the base of the covering winds around  the point at infinity once; the contour crosses only the branch cuts ending at $\infty_i$). These contours are also required to lie inside the fundamental polygon of the surface. 

\end{itemize}

\begin{proposition}
\label{prop_basis}
The contours $\{\lef_k(z)\}_{k=1}^n$ defined above on the covering  $\lambda:\surf\to\cp$ constitute a basis in the space $\Lambda^*(z).$
\end{proposition}
{\it Proof.}
The proof follows from the definition of the space $\Lambda^*(z)$ as the first relative homology group of the pair $(\surf^z,\d\surf^z),$ see Section \ref{sect_contours}. 

Alternatively, it is easy to see that  the contours $\c_k(z)$ for finite $z\neq 0$ constructed in Section \ref{sect_contours} can be decomposed in the space $\Lambda^*(z)$ into linear combinations of the contours $\lef_k(z)$ listed in the proposition. To every contour $\c_k(z)$ for $z\in\C\setminus \{0\}$ a linear combination of the paths $\{\calW_{\io j}(z)\}$ and $\{\calT_{i;\alpha}(z)\}$  can be added so that the result is a closed  contour on the surface not containing any of the points $\infty_i.$ Such a closed contour  is representable in the space $\Lambda^*(z)$ as a linear combination of the $a$- and $b$-cycles on the surface and the contours $\{\calV_i\}$ encircling the points $\{\infty_i\}.$
$\Box$ 

The next proposition gives a solution of system (\ref{RHz}), (\ref{RHi})  defined in the domain $D.$

\begin{proposition}
\label{thm_Psi0}
Let $D$ be a disc neighbourhood of $z=0$ with a branch cut chosen along the segment of the ray $l_-$ lying in the disc. 
The following matrix $\tilde{\Psi}_\io(z),$ defined for $z\in D,$ solves equations (\ref{RHz}) and (\ref{RHi}):
\beqn
\left(\tilde{\Psi}_\io(z)\right)_{ij}:=\frac{1}{2\i\sqrt{\pi}} \frac{1}{\sqrt{z}} \int_{\lef_j(z)} {\rm e}^{z\,\l(Q)}W(Q,P_i) .
\label{Psi0}
\eeqn
Here $\lef_j(z)$ are the contours from the basis in the space $\Lambda^*(z)$ given by Proposition \ref{prop_basis}.
\end{proposition}

{\it Proof.}  
 The integrals (\ref{Psi0}) do not depend on the direction in which contours of integration $\lef_j(z)$ approach points at infinity as long as  the argument ${\rm arg}\{z\l(Q)\}$  remains between $\pi/2$ and $3\pi/2$, as in (\ref{gammaz1}), for any point $Q$ on the contour in a neighbourhood of one of the points $\infty_i.$
Therefore, we can consider $\lef_j(z)$ as contours independent of $z$ when differentiating the matrix $\tilde{\Psi}_\io$ with respect to $z$. Their dependence on $z$ only comes into play when one studies the global behaviour of the solution, such as monodromy of $\tilde{\Psi}_\io$ as $z$ runs around the origin. 

Thus, the proof of the proposition repeats the first part of the proof of Theorem \ref{thm_Psi_rl}.
$\Box$

Let us now look at the monodromy transformation of the solution $\tilde{\Psi}_\io$ under analytic continuation around the origin. This transformation consists of the change of sign in the overall factor $1/\sqrt{z}$ in (\ref{Psi0}) and of the transformation of the integration contours $\lef_j(z)$. The next theorem gives the Jordan form of the monodromy matrix. 

\begin{theorem}
\label{thm_monodromy}
When the argument $z$ of the solution $\tilde{\Psi}_\io(z)$ (\ref{Psi0}) encircles the point $z=0$ in the counterclockwise direction, the matrix $\tilde{\Psi}_\io(z)$ transforms to another solution $\tilde{\Psi}_\io(z)\tilde{M}_\io$ with the monodromy matrix $\tilde{M}_\io$ having the following blocks in the Jordan canonical form:
\begin{itemize}
\item $2g$ blocks of the size $1$ with the eigenvalue $-1$
\item $m$ blocks of the size $2$ with the eigenvalue $-1$
\item $n_i$ blocks of the size $1$ with the eigenvalues $-{\rm e}^{2\pi \i  \alpha/(n_i+1)}$  where $\alpha = 1, \dots, n_i$ for any $i = 0, \dots, m$ 
\end{itemize}
\end{theorem}

\begin{remark}
\label{rmk_eigenvalues}
{\rm The eigenvalues of the monodromy matrix $\tilde{M}_\io$ are equal to ${\rm e}^{2\pi \i \mu_j}$, where $\{\mu_j\}$ are the eigenvalues of the matrix $V$, the spectrum (see Proposition \ref{prop_spectrum}) of the associated Frobenius manifold.    }
\end{remark}

{\it Proof.}
The quantity under the integral in (\ref{Psi0}) is singlevalued in $z$, but the contours of integration change as $z$ winds around the origin. 
Therefore, the monodromy matrix, up to the minus sign, is given by the matrix which transforms the system of cycles $\{\lef_j(z)\}$ to the new system  of similar cycles $\{\lef_j(z{\rm e}^{2\pi \i})\}$ obtained from $\{\lef_j(z)\}$ by a smooth deformation following the change in the argument of $z$. 

The cycles $\lef_j(z)$ for $j=1,\dots, 2g+m$ (the cycles $\{a_k;b_k\}$ and $\{\calV_i\}$) do not depend on $z$. 

For the remaining contours, $\lef_j(z), \;\; j>2g+m,$ the increase of $2\pi$ in the argument of $z$ 
results by (\ref{gammaz1}) in the turn of the end-parts of the contours by the angle of $2\pi/(n_i+1)$ in the local parameter (recall, that we take the local parameter at $\infty_i$ to be $\l^{-1/(n_i+1)}$).

Therefore, the path $\calW_{\io k}$ for $1\leq k\leq m$ connecting $\infty_\io$ to $\infty_k$ transforms to $\tilde{\calW}_{\io k}=\calW_{\io k} - \calT_{\io;\ione} + \calT_{k;\ione}$ if $n_\io>0$ and $n_k>0;$ if any of the ramification indices vanishes, the corresponding cycle $\calT_{\io;\ione}$ or $\calT_{k;\ione}$  is replaced by $\calV_\io$ or $\calV_k,$ respectively, in the above expression for the transformed contour $\tilde{\calW}_{\io k}.$ 

The contours $\{\calT_{i;\alpha}\}$  permute cyclically in each group of contours connecting one of $\infty_i$ with itself: for $i \neq 0$  the group 
 is $\{ \calT_{i;\ione}, \calT_{i;\itwo},\dots,\calT_{i;n_i},$ $ \calV_i - \calT_{i\ione}- \calT_{i\itwo} - \dots - \calT_{i;n_i}\}$; for $i=0$ 
 the cyclic permutation is done in the following group of contours: $\{ \calT_{\io;\ione}, \dots,\calT_{\io;n_0}, - \calT_{\io;\ione} - \dots - \calT_{\io;n_0} - \sum_{i=1}^m \calV_i \}.$

From these considerations one sees that for the first $2g+m$ columns of the monodromy matrix the only non-zero element is  $-1$ on the diagonal of the matrix. The minus sign comes from the factor $1/\sqrt{z}$ in the solution $\tilde{\Psi}_\io(z)$ (\ref{Psi0}).  As is also easy to see, the first $2g$ rows and  the $m$ rows from $(2g+m+1)$th to $(2g+2m)$th (corresponding to the paths connecting points at infinity on different sheets) have the similar structure: their only non-zero elements are $-1$ on the diagonal.  Thus, the monodromy matrix has $2g+2m$ (counting multiplicity) eigenvalues equal to $-1.$

In order to find its Jordan canonical form, we introduce a new basis of contours in the space $\Lambda^*(z),$ $z\in D,$ which will be referred to as the Jordan basis of contours.  Formula (\ref{Psi0}) with the contours $\gamma_j(z)$ replaced by the new contours defines a solution $\Psi_{\io}(z)$ in the domain $D$ (see formula (\ref{Psi0J})). The monodromy transformation of the solution $\Psi_{\io}(z)$ under analytic continuation around $z=0$ counterclockwise will be denoted by  ${M}_{\io};$ it is given by the matrix in the Jordan canonical form.

The Jordan basis of contours $\{\Gamma_j\}_{j=1}^n$ consists of:  

\begin{itemize}

\item $a-$ and $b-$ cycles: $\{a_k;\;b_k\}_{k=1}^g ;$

\item $\{\Upsilon_{2k-1}\}_{k=1}^m,$ where $\Upsilon_{2k-1} := N \calV_k;$
 
\item  $\{\Upsilon_{2k}(z)\}_{k =1}^m:$ For a sheet not containing the point $\infty_k$ there are $(n_k+1)$  paths starting at the point at infinity on the given sheet and ending at the point $\infty_k$ on one of the $(n_k+1)$ sheets glued together at the point $\infty_k.$  Let
the end-parts of the paths approach the points at infinity in the direction specified by (\ref{gammaz1}). The contour $\Upsilon_{2k}(z)$ is the sum of these paths oriented  from the point $\infty_k$ (the sum is taken over all sheets not containing $\infty_k$). As is easy to see, the contours $\Upsilon_{2k}(z)$ are linear combinations of the contours $\calW_{\io i}(z)$ from  Proposition \ref{prop_basis}.

\item $\{ \Delta_{i;\alpha}(z) \mid \alpha=1,\dots,n_i \}$ for each $i=0,\dots,m:$ 
\beqs
\Delta_{i;\alpha} (z):= \calV_i + \sum_{s = 1}^{n_i} (\varepsilon_{i;\alpha(n_i+1 - s)} - 1) \calT_{i;s}(z) , \hsp \mbox{where} \hsp \varepsilon_{i;k} = {\rm exp}{\frac{2\pi \i k}{n_i+1}} .
\eeqs
\end{itemize}

Let us now see how these contours transform when $z$ goes around zero. The $a$- and $b$-cycles and the cycles $\Upsilon_{2k-1}$ do not change.
Using the above definition of the contours $\Upsilon_{2k},$ after a simple computation we see that $\Upsilon_{2k}(z{\rm e}^{2\pi\i}) = \Upsilon_{2k}(z) - \Upsilon_{2k-1}(z).$ 
For the contours $\Delta_{i;\alpha}(z)$ using the relation $\varepsilon_{i;-\alpha} = \varepsilon_{i;\alpha n_i}$ for $\alpha<n_i$ we get $
\Delta_{i;\alpha}(z{\rm e}^{2\pi\i}) = \varepsilon_{i;\alpha} \Delta_{i;\alpha}(z) .$

Thus, we obtain a solution $\Psi_{\io}$ defined by 
\beqn
\left({\Psi}_\io(z)\right)_{ij}:=\frac{1}{2\i\sqrt{\pi}} \frac{1}{\sqrt{z}} \int_{\Gamma_j(z)} {\rm e}^{z\,\l(Q)}W(Q,P_i) 
\label{Psi0J}
\eeqn
with integration contours being $\{\Gamma_j(z)\}_{j=1}^n=\{\;\{a_k, b_k\}_{k=1}^g;\; \{\Upsilon_{2k-1}, \Upsilon_{2k}\}_{k=1}^m;\; \{\Delta_{i;\alpha} \mid \alpha = 1,\dots, n_i\}_{i=1}^m \;\}$ taken in this order. As can be seen from the above calculation, the mo\-nod\-romy matrix $M_\io$ of $\Psi_{\io}(z)$ at $z=0$ is in the Jordan form claimed in the theorem.  
$\Box$

Analogously to Theorem \ref{thm_Stokes} one finds the connection matrix $C$ defined by (\ref{CM}): it gives coordinates of the Jordan basis of contours $\{\Gamma_j(z)\}$  from Theorem \ref{thm_monodromy} with respect to the basis $\{\c^r_j\}_{j=1}^n$ in the space $\Lambda^*(z)$ with $z$ belonging to the intersection of the domain $D$ and the right half-plane.

\begin{proposition}
\label{thm_connection}
Consider a Hurwitz Frobenius manifold and the corresponding linear system (\ref{RHz}). Let $l = l_-\bigcup l_+$ be an admissible line in the $z$-plane for the system, which divides the plane into two half-planes $\Pi^r$ and $\Pi^l.$ 
Let the contours $\{\Gamma_j\}_{j=1}^n$ be the Jordan basis of contours constructed above, and the system of contours $\{\c^r_j\}$ be as in Section \ref{sect_infty}. Then the matrix $C$ such that the equality
\begin{equation*}
(\Gamma_1,\dots,\Gamma_n) = (\c_1^r, \dots, \c_n^r)C
\end{equation*}
holds in the space $\Lambda^*(z)$ with finite $z\in\Pi^r$ is the connection matrix (\ref{CM}).
\end{proposition}

Examples of computation of the connection matrix will be given in Section \ref{sect_examples}.

\subsection{Other Frobenius structures on Hurwitz spaces}
\label{sect_other}

\subsubsection{Deformed Frobenius manifolds}
\label{sect_deform}

By deformations of Hurwitz Frobenius manifolds we mean the Frobenius structures on Hurwitz spaces $\widehat{\M}_{g;n_\io,\dots,n_m}$ with rotation coefficients (\ref{Wqrot}) given by the bidifferential $\Wq$ (\ref{Wq-def}). We shall say that associated to these manifolds is the {\it deformed Riemann-Hilbert problem}. The deformed system of matrix differential equations (\ref{RHz}), (\ref{RHi}) has the form
\begin{eqnarray}
&&\d_z\Psi_\q(z) = (U+\frac{1}{z}V_\q)\Psi_\q(z) ,
\label{RHzq} \\\
&&\d_{\l_i}\Psi_\q (z) = (zE_i - [E_i,\Gamma_\q])\Psi_\q (z),
\label{RHiq}
\end{eqnarray}
 where  the diagonal matrix $U$ is the same as before and  $\Gamma_\q$ and $V_\q$ denote  the matrices $\Gamma$ and $V$ from (\ref{V}) built from rotation coefficients (\ref{Wqrot}). 

A solution $\Psi_\q$ of the deformed Riemann-Hilbert problem (\ref{RHzq}), (\ref{RHiq}) can be written similarly to the non-deformed case. Theorem \ref{thm_Psi_rl} with the bidifferential $W$ replaced by its deformation $\Wq$ (\ref{Wq-def}) gives a  solution to the system (\ref{RHzq}), (\ref{RHiq}) and, therefore, to the Riemann-Hilbert problem associated to the deformations of Hurwitz Frobenius manifolds:
\beqn
(\Psi^{l/r}_\q)_{ij} (z) :=\frac{1}{2\i\sqrt{\pi}}  \frac{1}{\sqrt{z}} \int_{\c^{l/r}_j} {\rm e}^{z\,\l(Q)}\Wq(Q,P_i) .
\label{Psiq_rl}
\eeqn
 A proposition similar to Proposition \ref{thm_Psi0} holds for the deformed problem. Namely, formula (\ref{Psi0}) with $W$ replaced by $\Wq$ gives a solution to (\ref{RHzq}), (\ref{RHiq}) defined in a the domain $D$ at the origin. 

The next theorem describes the relationship between solutions to the deformed and non-deformed systems.
\begin{theorem}
\label{thm_PsiPsiq}
The solution $\Psi$ of the system (\ref{RHz}), (\ref{RHi}) and the solution $\Psi_\q$ of the system (\ref{RHzq}), (\ref{RHiq}) are related by
\beqn
\Psi_\q (z) = \left(\Id -  \frac{1}{z} \Tq \right) \Psi (z) .
\label{PsiPsiq}
\eeqn
Here $\Id$ denotes the identity matrix; and  $\Psi$ denotes either of the solutions $\Psi^r$ or $\Psi^{l}$ given in their domains by Theorem \ref{thm_Psi_rl}. Similarly, $\Psi_\q$  denotes the respective deformed solution. The matrix $\Tq$ is a symmetric matrix with the entries:
\beqn
(\Tq)_{ij} = \pi \i \sum_{k,l = 1}^g \left(\B+\q \right)_{kl}^{-1} \omega_k(P_i)\omega_l(P_j) ,
\label{Tq}
\eeqn
where $\{\omega_k\}_{k=1}^g$ is the basis of holomorphic normalized differentials  and $\B$ is the Riemann matrix of the surface $\surf.$ The constant symmetric matrix $\q$ is the matrix of parameters from (\ref{Wq-def}).
\end{theorem}
{\it Proof.} The theorem can be proved by a direct computation as follows. Relation (\ref{PsiPsiq}) is equivalent to
\beqn
\label{comp1}
\int_{\rho}{\rm e}^{z\l(Q)}W_\q(Q,P_i) =\int_{\rho} {\rm e}^{z\l(Q)} \sum_{j=1}^n \left( \Id - \frac{1}{z}\Tq \right)_{ij} W(Q,P_j) ,
\eeqn
where $\rho$ stands for any of the contours $\lef_k(z),$ $\c^r_k$ or $\c^l_k$ from Proposition \ref{prop_basis} and Theorem \ref{thm_Psi_rl}.
Using the definition (\ref{Tq}) of the matrix $\Tq$ and the Rauch variational formula (\ref{Rauch}) for the holomorphic differentials $\omega_l,$ we rewrite the right-hand side of (\ref{comp1}) as follows:
\beqn
\label{comp2}
\int_{\rho}{\rm e}^{z\l(Q)}W(Q,P_i)  - \frac{2\pi\i}{z}  \sum_{k,l = 1}^g \left(\B+\q \right)_{kl}^{-1} \omega_k(P_i) \sum_{j=1}^n\d_{\l_j} \int_{\rho} {\rm e}^{z\l(Q)}  \omega_l(Q) .
\eeqn
For the sum of the derivatives with respect to the branch points $\l_j$  we have the relation: 
\beqn
\sum_{j=1}^n\d_{\l_j} \int_\rho {\rm e}^{z\,\l(P)} \omega_l(P)  = z\int_\rho {\rm e}^{z\,\l(P)}\omega_l(P) ,
\label{rel}
\eeqn
which can be proved by the method of the proof of Lemma \ref{elemma}, using the invariance of the holomorphic normalized differentials under biholomorphic mappings of Riemann surfaces. 

Plugging (\ref{rel}) into (\ref{comp2}) and using the definition (\ref{Wq-def}) of  the bidifferential $\Wq,$ we obtain the left-hand side of (\ref{comp1}).
$\Box$
\begin{corollary}
\label{cor_monodromy}
The matrices $S,{M}_\io, C$ for the deformed Riemann-Hilbert problem (\ref{RHzq}), (\ref{RHiq}) coincide with those for the non-deformed problem (\ref{RHz}), (\ref{RHi}) discussed in Sections \ref{sect_infty} and \ref{sect_origin}.
\end{corollary}

\begin{remark}
\label{rm_Tq}
{\rm 
The determinant of the matrix $G_\q(z):=\Id -   \Tq/{z}$  of the transformation (\ref{PsiPsiq}) equals $1$ as a corollary of the equality ${\rm det}\,\Psi^{r/l} = {\rm det}\,\Psi_\q^{r/l},$ which in turn follows from the fact that a theorem similar to Theorem \ref{thm_det} holds for the deformed solution $\Psi_\q^{r/l}.$ The matrix $G$ also satisfies
$G_\q^\iT(-z) G_\q(z) = \Id$ for the matrix ${\bf T}_{\q}^2$ vanishes due to the relation
\beqn
\label{tempTq}
\pi \i \sum_{j=1}^n \omega_k(P_j)\omega_l(P_j) \equiv \sum_{j=1}^n \d_{\l_j} \B_{kl} = 0.
\eeqn
The last equality in (\ref{tempTq}) is proved, for example, by putting $z=0$ and $\rho=b_k$ in formula (\ref{rel}). }
\end{remark}

\subsubsection{Real doubles of Frobenius manifolds}
\label{sect_doubles}

In this section we  work with the real Hurwitz space, i.e., with the space ${\M}_{g;n_0,\dots,n_m}$ of co\-ve\-rings where the set of local coordinates is formed by the branch points of the coverings and by their complex conjugates: $\{\l_j;\lb_j\}_{j=1}^n.$ Rotation coefficients of Frobenius structures on the real Hurwitz space \cite{doubles} are written (\ref{SBrot}) in terms of the Schiffer and Bergman kernels $\Omega$ and $B$ (\ref{Omegadef}), (\ref{Bdef}). A solution to the associated Riemann-Hilbert problem is a $2n\times 2n$ matrix which satisfies the linear system (\ref{RHz}), (\ref{RHi}) with the diagonal matrix $U = diag(\l_1,\dots,\l_n,\lb_1,\dots,\lb_n)$ and $2n \times 2n$ matrices $\Gamma$ and $V$ formed by the rotation coefficients (\ref{SBrot}) as described by (\ref{V}) and in Section \ref{sect_HF}. We shall refer to this Riemann-Hilbert problem as the Riemann-Hilbert problem for doubles. A solution to this problem can be written in terms of the Schiffer and Bergman kernels analogously to the formulas (\ref{Psi0}) and (\ref{Psi_rl}) for the matrix $\Psi$. The solution will be denoted by $\PsiSB$; it is given by the next theorem. 
\begin{theorem}
\label{thm_PsiSB}
The solution ${\Psi}^r_{\iOmega\iB}$ to the Riemann-Hilbert problem corresponding to the real doubles of Hurwitz Frobenius manifolds is the following $2n\times 2n$ matrix consisting of four $n\times n$ blocks:
\beqn
\label{PsiSB_r}
{\Psi}^r_{\iOmega\iB}(z) = \frac{1}{2\i\sqrt{\pi}} \frac{1}{\sqrt{z}} \left( \begin{array}{cc} \left( \int_{\c_j^r} {\rm e}^{z\,\l(Q)}\Omega(Q,P_i) \right) & \left( \int_{\c_j^r} {\rm e}^{z\,\overline{\l(Q)}} B(\Qbar,P_i) \right) \\ \\\left( \int_{\c_j^r} {\rm e}^{z\,\l(Q)} B (Q,\Pbar_i) \right) & \left( \int_{\c_j^r} {\rm e}^{z\,\overline{\l(Q)}\;} \overline{\Omega(Q,P_i)} \right) \end{array}\right) .
\eeqn
 Each block is given by its $(ij)$-entry; $i,j\in\{1,\dots,n\}.$ The rows in blocks are labeled by ramification points, i.e., by the indices $i\in \{1,\dots,n\}$.
The contours $\c_j^r$ are the same as in Theorem \ref{thm_Psi_rl}.
The solution ${\Psi}^l_{\iOmega\iB}$ in the left half-plane $\Pi^l$ is obtained from (\ref{PsiSB_r}) by replacing the contours $\c_j^r$ with the contours  $\c^{l}_j.$ 
\end{theorem}

Proof of the theorem is analogous to that given above for Theorem \ref{thm_Psi_rl}.
Similarly to Theorem \ref{thm_det} we compute the determinant of the solution $\PsiSB(z).$
\begin{theorem}
The determinant of the solution ${\Psi}^{l/r}_{\iOmega\iB}$ from Theorem \ref{thm_PsiSB} is given by:
\beqs
{\rm det} \; {\Psi}^{l/r}_{\iOmega\iB}(z) = {\rm exp} \{z\sum_{k=1}^n (\lambda_k+\bar{\lambda}_k)\} .
\eeqs

\end{theorem}

The next theorem establishes a relationship between the solution $\Psi(z)$ to the Riemann-Hilbert problem for the family of Hurwitz Frobenius manifolds and the solution $\PsiSB (z)$ to the Riemann-Hilbert problem for their real doubles. 

\begin{theorem}
\label{thm_doubles}
Let $\Psi(z)$ be the $n\times n$ matrix solution to the Riemann-Hilbert problem from Theorem \ref{thm_Psi_rl} (expressed in terms of the bidifferential $W$).   The solution $\PsiSB$ from Theorem \ref{thm_PsiSB} of the Riemann-Hilbert problem for the real doubles can be obtained from the matrix function $\Psi$ by the following transformation: 
\beqn
\PsiSB (z) = \left(\Id -  \frac{1}{z} \Tdoubles \right) \left(\begin{array}{cc}\Psi(z) & 0 \\0 &\overline{\Psi(\bar{z})} \end{array}\right) .
\label{PsiPsiSB}
\eeqn
Here $\Id$ denotes the identity matrix. By $\Psi$ we denote any of the matrices  $\Psi^r,$ $\Psi^l$ considered in their domains; and $\PsiSB$ stands for one of the  $\Psi_{\iOmega\iB}^r,$ $\Psi_{\iOmega\iB}^l,$ respectively. 
The matrix $\Tdoubles$ is the following symmetric $2n\times 2n$ matrix consisting of the four $n\times n$ blocks. Each block is given by its $(ij)$-entry; $i,j\in\{1,\dots,n\}:$
\beqn
\Tdoubles := \frac{\pi}{2}     \left( \begin{array}{cc} \left( \sum_{k,l = 1}^g \left({\rm Im} \B \right)_{kl}^{-1}  \omega_k(P_i)\omega_l(P_j)  \right) & \left( - \sum_{k,l = 1}^g \left({\rm Im} \B \right)_{kl}^{-1}  \omega_k(P_i) \overline{\omega_l(P_j)}  \right) \\
 \\ \left( - \sum_{k,l = 1}^g \left({\rm Im} \B \right)_{kl}^{-1} \overline{\omega_k(P_i)}\omega_l(P_j)  \right) & \left( \sum_{k,l = 1}^g \left({\rm Im} \B \right)_{kl}^{-1}  \overline{\omega_k(P_i)} \;\overline{\omega_l(P_j)}  \right) \end{array}\right) ,
\label{Tdoubles}
\eeqn
where $\{\omega_k\}_{k=1}^g$ is the basis of holomorphic normalized differentials  and $\B$ is the Riemann matrix of the surface $\surf.$ 
\end{theorem}
The proof can be obtained by a direct calculation analogously to the proof of Theorem \ref{thm_PsiPsiq}.
\begin{corollary}
\label{cor_doubleStokes}
The Stokes matrix, the monodromy  at the origin and  and the connection matrices corresponding to the Riemann-Hilbert problem for the real doubles have a block-diagonal structure with two blocks. The blocks are given by the respective matrices from the monodromy data of  the Riemann-Hilbert problem for Frobenius structures on the complex Hurwitz space (described in Sections \ref{sect_infty}  and \ref{sect_origin}).
\end{corollary}
\begin{remark}
{\rm Similarly to the matrix ${\bf T}_{\q}^2,$ the matrix $\Tdoubles^2$ (\ref{Tdoubles}) vanishes as a corollary of relation (\ref{tempTq}). Therefore, similarly to $G_\q$ in Remark \ref{rm_Tq}, the  matrix $G(z):=\Id -   \Tdoubles/z$ of the transformation (\ref{PsiPsiSB}) satisfies the relations:  
${\rm det}\, G(z) = 1;$ $ G^\iT(-z) G(z) = \Id . $
}\end{remark}

\subsection{Isomonodromic tau-function}

The isomonodromic tau-function $\tau_\iI$ associated to monodromy preserving deformations of a system of linear ordinary differential equations was introduced in \cite{JMU} . The function $\tau_\iI$ is a function of deformation parameters; it plays an important role in the theory of isomonodromic deformations. In \cite{JMU} it was conjectured that $\tau_\iI$ is holomorphic everywhere in the space of deformation parameters outside of the hyperplanes where the values of any two deformation parameters coincide. This conjecture was proved in \cite{Bolibruch, Malgrange}. In our case the deformation parameters are the coordinates $\{\l_k\}$ on the Frobenius manifold. The tau-function $\tau_\iI$ is thus holomorphic on the universal covering of the space ${\mathbb C}^n\setminus \{(\l_\ione,\dots,\l_n)\mid \l_k=\l_l {\mbox{ with }} k\neq l\}.$  The set of zeros of the function $\tau_\iI$ in this space is called  {\it the Malgrange divisor}; the solvability of the corresponding  Riemann-Hilbert problem can be described in terms of this divisor. For example, for the given matrices $\mu,$ $R$ and $S$ (see Section \ref{sect_formulation}) the solution always exists outside of the Malgrange divisor \cite{BolibruchOrders}.

In this section we summarize the known results on isomonodromic tau-functions of the systems (\ref{RHz}), (\ref{RHi}) associated to Frobenius manifolds studied in the paper: the tau-functions are computed in terms of objects defined on the underlying Riemann surface. 

For our system (\ref{RHz}), (\ref{RHi}) corresponding to a Frobenius manifold, the definition \cite{JMU} of the tau-function reduces,  according to Examples 5.2 and 5.3 in \cite{JMU}, to:
\beqn
\frac{\d \log \tau_\iI}{\d \l_i} :=  - \sum_{j \neq i, j = 1} ^n \beta_{ij}^2 (\l_i - \l_j) 
, \qquad i = 1, \dots, n .
\label{tauiso}
\eeqn
As was shown in \cite{KokKorG}, for the Frobenius manifolds with rotation coefficients $ \beta_{ij}$ given by the bi\-dif\-fe\-ren\-tial $W$ (\ref{W-def}), 
the isomonodromic tau-function $\tau_\iI$ (\ref{tauiso})  coincides with the so-called {\it Bergman tau-function}  $\tau_\iW,$ introduced and computed in \cite{KokKorB} (note that the definition  of $\log\tau_\iI$ used in \cite{KokKorG}
differs by a factor of $-1/2$ from the definition of \cite{JMU} cited here (\ref{tauiso})). 
The Bergman tau-function is defined in terms of the bidifferential $W$ as follows. 

Denote by $S^\iW$ the following term in the asymptotics of  $W(P,Q)$ (\ref{W-def}) near the diagonal $ P \sim Q :$
\beqs
W(P,Q) \underset{Q \sim P}{=} \left( \frac{1}{(x(P) - x(Q))^2} + S^\iW(x(P)) + o(1) \right) dx(P) dx(Q)
\eeqs
(the quantity $6S^\iW(x(P))$ is called the Bergman projective connection \cite{Fay92}). Choosing the local parameter to be $x_i(P)  = \sqrt{ \l - \l_i }$ we denote by $S^\iW_i$ the value of $S^\iW$ at a ramification point $P_i:$
\beqn
S^\iW_i = S^\iW(x_i)\Big|_{x_i = 0} .
\label{SWi}
\eeqn
Since the singular part of $W$-kernel in a neighbourhood of the point $P_i$ does not depend on coordinates $\{ \l_j \},$ the Rauch variational formulas (\ref{WRauch}) imply ${\d_{\l_j} S^\iW_i} =  W^2(P_i,P_j)/2 .$
The symmetry of this expression with respect to the indices $i$ and $j$ provides the compatibility of the system of differential equations which defines the tau-function $\tau_\iW:$
\beqn
\frac{\d \log \tau_\iW }{\d \l_i} = - \frac{1}{2} S^\iW_i , \qquad  i=1, \dots, n  .
\label{tau-W}
\eeqn

The Bergman tau-function (\ref{tau-W}) first appeared in \cite{DimaRH} where it entered the expression for the isomonodromic tau-function corresponding to the Riemann-Hilbert problem with Fuchsian singularities and quasipermutational monodromy matrices. In \cite{KokKorB} $\tau_\iW$ was computed in terms of the prime form, theta-functions and holomorphic normalized differentials on the Riemann surface. It was also shown that $\tau_\iW$ has no zeros in the universal covering of the space ${\mathbb C}^n\setminus \{(\l_\ione,\dots,\l_n) \mid \l_k=\l_l {\mbox{ with }} k\neq l\}.$ This implies that the Malgrange divisor for the tau-function $\tau_\iI$ (\ref{tauiso}) is empty. Hence the Riemann-Hilbert problem associated to the Frobenius structures on Hurwitz spaces from \cite{2D} is solvable for any point of  the Hurwitz space $\widehat{\M}_{g;n_\io,\dots,n_m}.$

The isomonodromic tau-function  $\tau_{\iI\q}$ for the deformed Frobenius manifold structures can be also expressed in terms of the Bergman tau-function. As was shown in \cite{deform}, $\tau_{\iI\q} = \tau_\iW \det(\B+\q)$ (note that the definition of ${\rm log}\tau_\iI$ in \cite{deform} is consistent with that of \cite{KokKorG} and therefore is different from the one used here by a factor of $-1/2$).

In the case of real doubles, the Frobenius manifolds with rotation coefficients (\ref{SBrot}), let us denote the isomonodromic tau-function by $\tau_\iI^{\iOmega\iB};$ the definition (\ref{tauiso}) in this case  becomes:
\begin{align}
\begin{split}
\frac{\d \log \tau_{\iI}^{\iOmega\iB}}{\d \l_i} = - \sum _{j \neq i, j=1}^L \beta_{ij}^2 (\l_i - \l_j) - \sum _{j=1}^L \beta_{i\bar{j}}^2 (\l_i - \bar{\lambda}_j) ,  \\
\frac{\d \log \tau_\iI^{\iOmega\iB}}{\d \bar{\l}_i} = - \sum _{j=1}^L \beta_{\bar{i}j}^2 (\lb_i - \l_j) - \sum _{j \neq i, j=1}^L \beta_{\bar{i}\bar{j}}^2 (\bar{\l}_i - \bar{\l}_j) .
\label{tauiso2}
\end{split}
\end{align}
The tau-function $\tau_{\iI}^{\iOmega\iB}$ is related \cite{doubles} to  the Bergman tau-function by $\tau_{\iI}^{\iOmega\iB} =  | \tau_\iW |^2 \det ({\rm Im} \B).$

Thus, we can formulate the following
\begin{theorem}
\label{thm_tau}
The isomonodromic tau-function $\tau_\iI$ (\ref{tauiso}) with $\beta_{ij} = W(P_i,P_j)/2$ (corresponding to the Frobenius structures on Hurwitz spaces  from \cite{2D})  coincides with the so-called Bergman tau-function $\tau_\iW$ (\ref{tau-W}) computed in \cite{KokKorB}: $\tau_\iI = \tau_\iW .$

The isomonodromic tau-function defined by (\ref{tauiso}) with $\beta_{ij} = \Wq(P_i,P_j)/2$ (corresponding to the deformations of the Frobenius structures on Hurwitz spaces from \cite{2D}) is given by 
\beqs
\tau_{\iI\q} = \tau_\iW \det(\B+\q) ,
\eeqs
where $\B$ is the Riemann matrix of the underlying surface and $\q$ is the symmetric matrix of parameters.

The isomonodromic tau-function defined by (\ref{tauiso2}) and the rotation coefficients (\ref{SBrot}) corresponding to the real doubles of the Frobenius structures on Hurwitz spaces from \cite{2D} is given by 
\beqs
\tau_{\iI}^{\iOmega\iB} = | \tau_\iW |^2 \det ({\rm Im} \B) . 
\eeqs
\end{theorem}

Since the Bergman tau-function $\tau_\iW$ vanishes nowhere  in the universal covering of the space ${\mathbb C}^n\setminus \{(\l_\ione,\dots,\l_n) \mid \l_k=\l_l {\mbox{ with }} k\neq l\},$
Theorem \ref{thm_tau} implies that the Malgrange divisor of the isomonodromic tau-function $\tau_{\iI\q}$ coincides with the divisor (\ref{divisor}) in the space $\widehat{\M}_{g;n_\io,\dots,n_m}.$ Formula (\ref{Psiq_rl}) shows that, in accordance with the general theory, the solution $\Psi_\q$ to the Riemann-Hilbert problem corresponding to the deformations of Frobenius manifolds fails to exist exactly at the points of the divisor (\ref{divisor}). 
The Malgrange divisor of the tau-function $\tau_{\iI}^{\iOmega\iB}$ of the real doubles is empty.

\section{Monodromy data: examples}  
\label{sect_examples}

In this section we look at various examples of Hurwitz spaces and compute the Stokes matrix, the monodromy matrix at zero and the connection matrix 
 for the corresponding Riemann-Hilbert problems. We restrict ourselves to the case of rotation coefficients (\ref{Wrot}) given by the bidifferential $W$ (i.e., to the case of Frobenius manifold structures on Hurwitz spaces from \cite{2D}). All written in this section is valid for the case of rotation coefficients given by the deformed bidifferential $\Wq;$ the monodromy matrices ($S,$ $C,$ $M$) for the  real doubles can be obtained from the ones computed here using Corollary \ref{cor_doubleStokes}, Section \ref{sect_doubles}.

\subsection{Set-up}
\label{sect_assumptions}

In this section we list a few assumptions on the coverings under which we work in the sequel. 

Recall that for a fixed line $l$ dividing the $z$-plane into the half-planes $\Pi^r$ and $\Pi^l$ the form of the Stokes matrix depends on the arrangement of the branch points on the plane:  an off-diagonal entry $S_{ij}$ vanishes if the corresponding Stokes ray $r_{ij}$ (\ref{rays}) belongs to the right half-plane $\Pi^r,$ or, in other words, if  ${\rm Re}\{z(\l_i-\l_j)\}>0$ for $z\in l_+.$ 

\begin{figure}[htb]
\centering
\subfigure{\includegraphics[width=5cm]{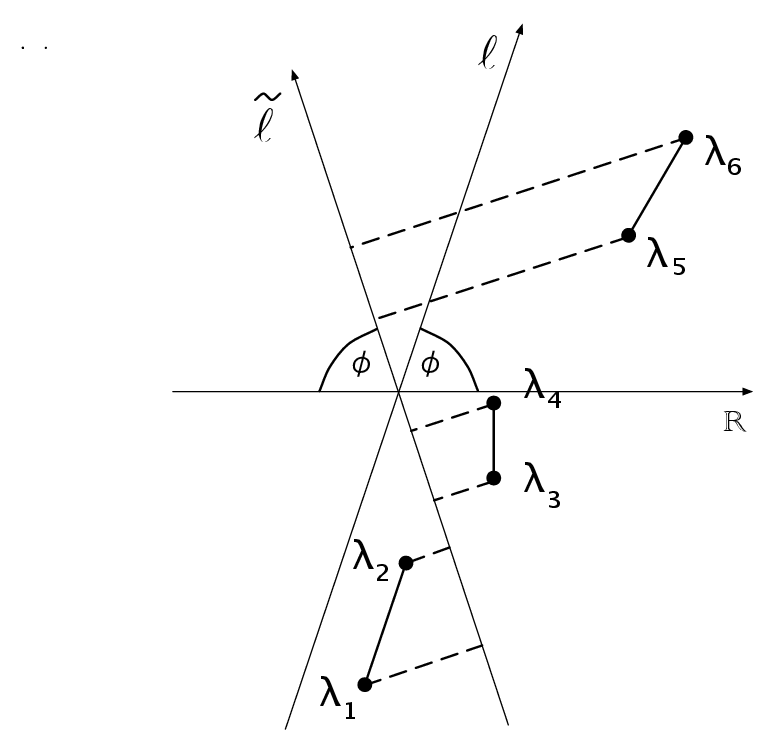}}
\caption{Ordering of the branch points.}
\label{fig_order}
\end{figure}

We shall work with the configurations of branch points for which the Stokes matrix is lower-triangular. Let $l$ be an oriented admissible line in the $z$-plane; and let $\phi$ be the angle between $l$ and the real axis.
Since the condition $S_{ij}=0$ for $i<j$ is equivalent to  ${\rm Re}\{z\l_i\}>{\rm Re}\{z\l_j\}$ for ${\rm arg} z = \phi$ and $i<j,$ we need  the branch points to be ordered so that after  rotating the set $\{\l_k\}_{k=1}^n$ by the angle $\phi$ we would get a set of points ordered according to the descending real part.
Such an ordering is shown in Figure \ref{fig_order}. 

Namely, we define an oriented line $\tilde{l}$ such that the angle between $\tilde{l}$ and the real line is $\pi-\phi.$ Then we
 place the branch points $\{\l_k\}_{k=1}^n$  on the $z$-plane and consider their orthogonal projections on the line $\tilde{l}.$ We enumerate the projections in the ascending order in the direction of the orientation of the line $\tilde{l}.$  Note that no two projections coincide since $l$ is an admissible line for the equation.  A branch point is then assigned the label of its projection on the line $\tilde{l}.$ For such ordering of the branch points, the entries of the Stokes matrix which lie above the diagonal vanish. Recall, that all diagonal entries of the matrix $S$ are equal to $1.$

In addition we assume that  
the branch cuts can be chosen to join the neighbouring ramification points, i.e.,  the points 
$P_1$ and $P_2$, $P_3$ and $P_4,$ and so on (where $\l_i = \l(P_i)$).

 Let us also assume the sheets of the covering to be ordered.  Recall that the contours  $\c_k^r$ and $\c_k^l$ are defined to pass through a neighbourhood of the corresponding ramification point $P_k$ in a determined way: see inequality (\ref{gammax}) and the comment after it. 
Therefore, we need to make the following assumption for the local parameters near the points $P_k.$ Among the two sheets glued together at the point $P_k$ we shall call the sheet with a smaller label the ``lower" sheet and the other one the ``upper" sheet. Then we assume that the branches of the local parameter $x_k(P)=\sqrt{\l(P)-\l_k}$ near the point $P_k$ are chosen so that the ray ${\rm arg} \, x_k  = 3\pi/4 - \phi/2$  belongs to the lower sheet and the ray ${\rm arg} \, x_k  = -\pi/4-\phi/2$   to the upper sheet glued together at the point $P_k.$ In other words, the contours $\c_k^r$ in a neighbourhood of the point $P_k$ pass from the lower to the upper sheet of the covering. 

Let us summarize the assumptions we made. In what follows we shall refer to them as assumptions A1, A2 and A3: 
\begin{itemize}

\item[A1:] The configuration of branch points is such that ${\rm Re}\{z\l_i\}>{\rm Re}\{z\l_j\}$ for ${\rm arg}\, z = \phi$ and $i<j,$ i.e., 
the branch points are ordered as shown in Figure \ref{fig_order}. Note that this ordering depends on the choice of the separating line $l.$ 

\item[A2:]
Pairs of ramification points with successive labels, i.e $P_1,$ $P_2,$ and  $P_3,$ $P_4,$ and so on are connected by a branch cut. 

\item[A3:] The contours $\c^r_k$  pass from the lower to the upper sheet of the covering in a neighbourhood of the ramification point $P_k.$

\end{itemize}

In the subsequent examples we assume that the separating line and the branch points are chosen as in Figure \ref{fig_order}.

\subsection{The Hurwitz space $\M_{\io;\io,\io}$}
\label{sect_H000}
\begin{figure}[htb]
\centering
\subfigure{\includegraphics[width=6cm]{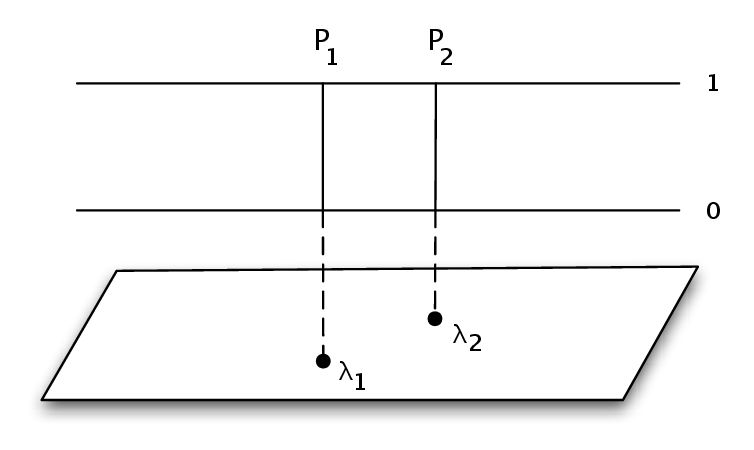}}
\caption{The Hurwitz diagram for the space $\M_{\io;\io,\io}.$ }
\label{fig_diagram000}
\end{figure}

The Hurwitz space $\M_{\io;\io,\io}$ is the space of two-fold genus zero coverings with two simple finite branch points $\l_1$ and $\l_2.$ These coverings can be graphically represented by the  Hurwitz diagram given by Figure \ref{fig_diagram000}.

Despite its simplicity, the monodromy data in this example exhibits a structure which is shared by the monodromy data of Frobenius structures on all   Hurwitz spaces considered below. Namely, we shall see that the Stokes matrix computed in this section will appear as a diagonal block in all subsequent examples of the Stokes matrix.

\begin{figure}[htb]
\centering
\qquad\qquad\qquad{\includegraphics[width=4cm]{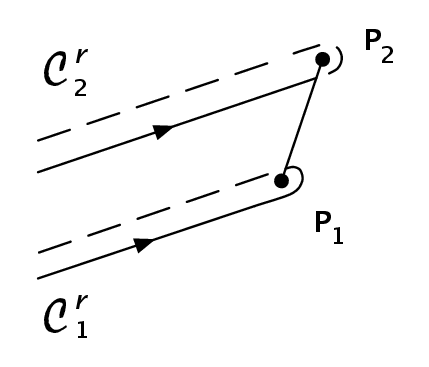}\qquad\subfigure{\includegraphics[width=7cm]{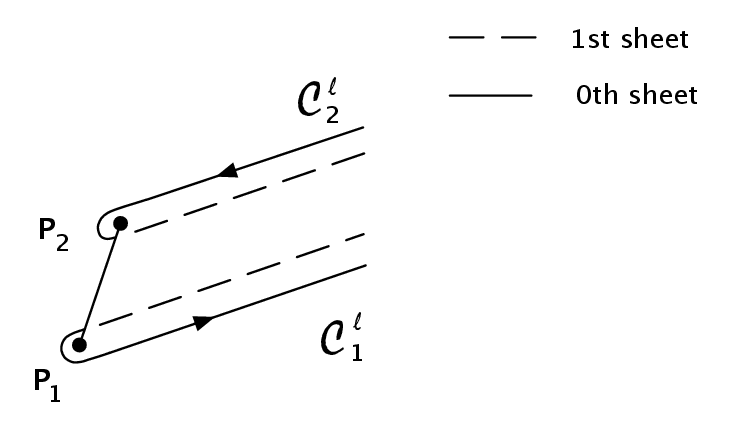}}}
\caption{Contours $\c_k^l$ and $\c_k^r$ for the space $\M_{\io;\io,\io}.$}
\label{fig_lr02}
\end{figure}

Let us fix a value  $z_+$ belonging to the ray $l_+.$ The solutions $\Psi^r$ and $\Psi^l$ (\ref{Psi_rl}) can be ana\-ly\-ti\-cally continued into a sector neighbourhood of the ray $l_+$ as in Proposition \ref{prop_cont} and Theorem \ref{thm_Stokes}. The corresponding contours $\hat{\c}_k^r$ and $\hat{\c}_k^l$ (see Proposition \ref{prop_cont}) belong to the space $\Lambda^*(z_+).$ Their projections on the base of the covering satisfy: 
\beqs
{\rm arg}(\zeta-\l_k) = \frac{3\pi}{2} - \phi -\epsilon  , \hsp \zeta \in pr(\hat{\c}_k^r) ; 
\eeqs
\beqs
{\rm arg}(\zeta-\l_k) = \frac{\pi}{2} - \phi +\epsilon , \hsp \zeta \in pr(\hat{\c}_k^l)
\eeqs
with  a small $\epsilon>0.$ A clockwise rotation of the projection $pr(\hat{\c}_k^r)$ towards $pr(\hat{\c}_k^l)$ in the $\zeta$-plane induces a deformation, smooth in a neighbourhood of the ramification point $P_k,$ which takes the contour $\hat{\c}_k^r$ into $\hat{\c}_k^l.$
Figure \ref{fig_lr02} shows the contours ${\c}_k^r$ and ${\c}_k^l,$ $k=1,2$ for the choice of the separating line $l$ as in Figure \ref{fig_order} (when drawing a picture we do not distinguish between the deformed contours $\hat{\c}_k^r,$ $\hat{\c}_k^l$ and the contours ${\c}_k^r,$ ${\c}_k^l$). 

The Stokes matrix gives the coordinates of the contours $\hat{\c}_k^l$ with respect to the basis of contours in the space $\Lambda^*(z_+),$ $z_+\in l_+,$ given by $\{\hat{\c}_k^r\},$ see Theorem \ref{thm_Stokes}.

The contours are related by: 
\beqs
\hat{\c}_1^l = \hat{\c}_1^r - 2\hat{\c}_2^r , \qquad
\hat{\c}_2^l = \hat{\c}_2^r.
\eeqs
This can be seen from the picture as follows. The contour $\hat{\c}_1^r-\hat{\c}_2^r$ is equivalent to a contour on the $0$th sheet encircling the branch cut counterclockwise (and to a contour on the first sheet encircling the branch cut clockwise). The same contour is equivalent to $\hat{\c}_1^l+\hat{\c}_2^l.$ Recall that the contours $\hat{\c}_k^{r/l}$ can go around the ramification point $P_k$ in either direction since the residues of the differentials (\ref{diffspace}) at the ramification points vanish. 

Thus the Stokes matrix of the Frobenius manifold structure on the Hurwitz space $\M_{\io;\io,\io}$ (i.e., the Stokes matrix (\ref{SM}) of the corresponding equation (\ref{RHz})) has the form: 
\beqn
\label{S000}
S^{\io;\io,\io} =   \left(\begin{array}{rr}
      1 & \;\;0  \vspace{.2cm} \\  
      -2 & \;\;1  \\ 
   \end{array} \right) .
\eeqn

Let us now compute the connection matrix (\ref{CM}). 

The contours $\{\c^r_\ione, \c^r_\itwo\}$ give a basis in the space
$\Lambda^*(z)$ for $z\in\Pi^r,$ $|z|<\infty.$ Another basis in this space is given by Proposition \ref{prop_basis}. It consists of the contour $\calV_\ione,$ encircling the point $\infty_\ione$ (the point on the sheet number $1$ projecting to $\l=\infty$ on the base) counterclockwise, and the contour $\calW_{\io \ione}$ going from the point $\infty_\io$ to $\infty_\ione$ and satisfying condition (\ref{gammaz1}) near the end points.  
For our covering these contours are: $\calW_{\io \ione} = {\c}_\itwo^r$,  $\calV_\ione= {\c}_\ione^r - {\c}_\itwo^r$ (we may as well take $\calW_{\io \ione} = {\c}_\ione^r$,  $\calV_\ione= {\c}_\ione^r - {\c}_\itwo^r$). 
The Jordan basis of contours (see proof of Theorem \ref{thm_monodromy}) then is $\Gamma_\ione=2{\calV_\ione}=2{\c}_\ione^r - 2{\c}_\itwo^r$ and $\Gamma_\itwo=-{\calW_{\io \ione}}=- {\c}_\itwo^r.$

The connection matrix $C$ (\ref{CM}) gives the coordinates of the contours $\{\Gamma_\ione,\,\Gamma_\itwo\}$ with respect to the basis  $\{\c^r_\ione,\c^r_\itwo\},$ see Theorem \ref{thm_Stokes}:
\beqs
C^{\io;\io,\io}  =   \left(\begin{array}{rr}
      2 & 0 \vspace{.2cm} \\  
      -2 & -1 \vspace{.2cm} \\ 
   \end{array} \right) .
\eeqs
Using the definition of the contours $\calV_\ione$ and $\calW_{\io \ione},$ we compute the monodromy matrix $\tilde{M}_\io$ for the solution $\tilde{\Psi}_\io$ (\ref{Psi0})  at the origin,
\beqs
\tilde{M}_\io^{\io;\io,\io} =   \left(\begin{array}{rr}
      -1 & -2  \vspace{.2cm} \\  
      0 & -1   \vspace{.2cm} \\ 
   \end{array} \right) ,
\eeqs
and see that its Jordan canonical form $M_{\io}^{\io;\io,\io}$ coincides with that given by Theorem \ref{thm_monodromy}.
The matrix $M_{\io}^{\io;\io,\io}$ can also be obtained from the Stokes matrix and the connection matrix using the relation $M_{\io} = C^{-1}S^\iT S^{-1} C.$

\subsection{The Hurwitz space $\M_{\io;\io,\io,\io}$} 
\label{sect_H0000}
\begin{figure}[htb]
\centering
\subfigure{\includegraphics[width=6cm]{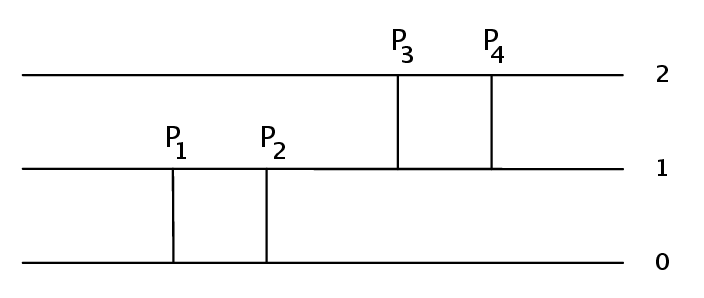}}
\caption{The Hurwitz diagram for the space $\M_{\io;\io,\io,\io}.$ }
\label{fig_diagram0000}
\end{figure}

The Hurwitz space $\M_{\io;\io,\io,\io}$ is the space of genus zero three-fold coverings with four simple finite ramification points.
Let us assume that the sheets are ordered as shown by the Hurwitz diagram in Figure \ref{fig_diagram0000}.

This example shows the construction of the Stokes matrix corresponding to the space of coverings obtained from the two-fold genus zero coverings by attaching an extra sheet along the branch cut connecting two additional branch points (under the assumption that the ordering in the resulting set of four branch points is as shown in Figure \ref{fig_order}). We shall see (formula (\ref{blockA})) how the $4\times 4$ Stokes matrix corresponding to the space $\M_{\io;\io,\io,\io}$ is related to the $2\times 2$ Stokes matrix computed in the previous section for the space $\M_{\io;\io,\io}$ of two-fold genus zero coverings. This will give us an insight into the way to find Stokes matrices for the Hurwitz spaces of coverings which can be obtained from the given one by attaching one extra sheet along a branch cut.

\begin{figure}[htb]
\centering
\qquad\qquad\qquad{\includegraphics[width=4.5cm]{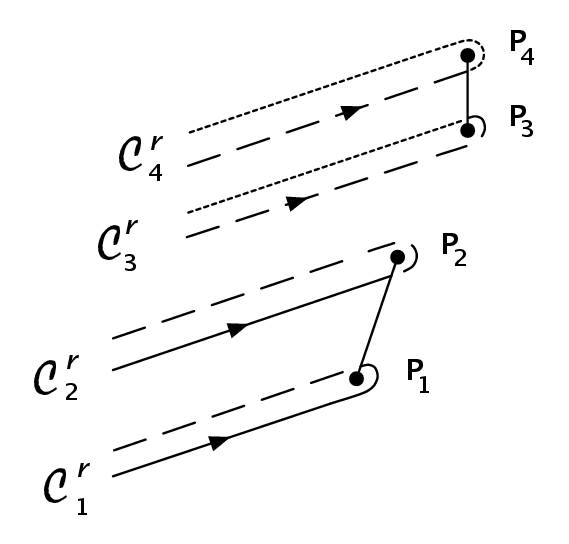}\qquad\subfigure{\includegraphics[width=7.5cm]{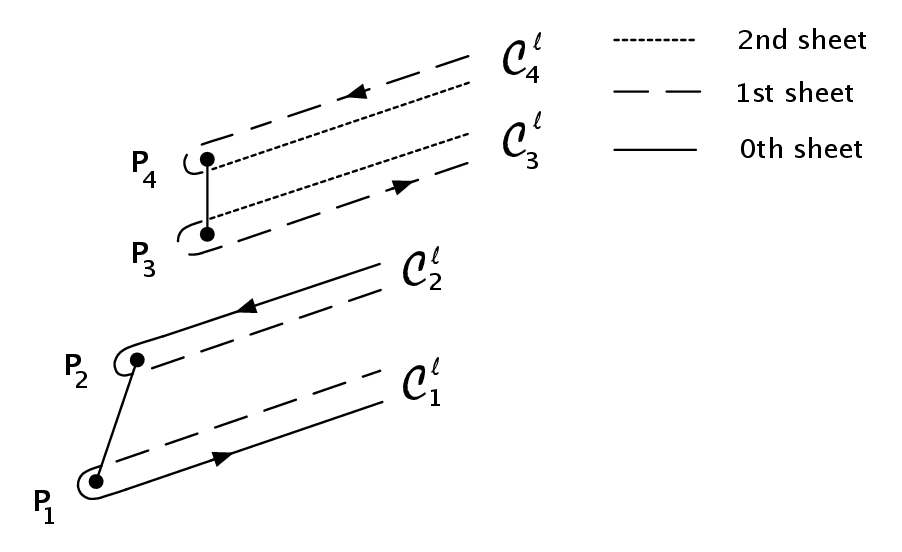}}}
\caption{Contours $\c_k^l$ and $\c_k^r$ for the space $\M_{\io;\io,\io,\io}.$}
\label{fig_lr04}
\end{figure}

As before, to compute the Stokes matrix  we need to find the transformation which takes the contours $\{\hat{\c}^r_k\}_{k=1}^4$ to the contours $\{\hat{\c}^l_k\}_{k=1}^4$ in the space $\Lambda^*(z_+)$ for $z_+\in l_+.$ The contours $\c^r_k$ and $\c^l_k$ are  shown in Figure \ref{fig_lr04}. 
Deforming them by a small rotation about $P_k$ towards each other, so that the contours $\c^r_k$ rotate clockwise, we obtain the contours $\hat{\c}^r_k$ and $\hat{\c}_k^l.$ In the space $\Lambda^*(z_+)$, we have the relations:
\beqs
\hat{\c}_1^l = \hat{\c}_1^r - 2\hat{\c}_2^r - \hat{\c}_3^r +\hat{\c}_4^r, \qquad \hat{\c}_2^l= \hat{\c}_2^r+\hat{\c}_3^r -\hat{\c}_4^r,\qquad \hat{\c}_3^l=\hat{\c}_3^r-2\hat{\c}_4^r,\qquad \hat{\c}_4^l=\hat{\c}_4^r.
\eeqs
To verify the relations we note that the contour $\hat{\c}_1^r-\hat{\c}_2^r$ is equivalent to the contour encircling the cut $[P_1,P_2]$ counterclockwise on the $0$th sheet; the same contour can be obtained as $\hat{\c}_1^l+\hat{\c}_2^l.$ The contour encircling the same branch cut on the $1$st sheet is given by $\hat{\c}_2^r-\hat{\c}_1^r.$ The contour $\hat{\c}_3^r-\hat{\c}_4^r=\hat{\c}_3^l+\hat{\c}_4^l$ encircles the branch cut $[P_3,P_4]$ counterclockwise on the $1$st sheet. 

Therefore, the Stokes matrix (\ref{SM}) of the Frobenius manifold structure on the Hurwitz space $\M_{\io;\io,\io,\io}$ is given by
\beqs
S^{\io;\io,\io,\io} =   \left(\begin{array}{rrrr}
      1 & 0 & 0 & \;\;0   \vspace{.2cm} \\  
      -2 & 1 & 0 &\;\; 0   \vspace{.2cm} \\  
      -1 & 1 & 1 & \;\;0   \vspace{.2cm} \\  
      1 & -1 & -2 & \;\;1  \\ 
   \end{array} \right) .
\eeqs

It is convenient to rewrite this matrix using the Stokes matrix $S^{\io;\io,\io}$ (\ref{S000}) from the previous section: 
\beqn
\label{blockA}
S^{\io;\io,\io,\io} =   \left(\begin{array}{cc}
      S^{\io;\io,\io} & \;\;{\bf 0}   \vspace{.2cm} \\  
      {\bf A} & S^{\io;\io,\io}   \\  
   \end{array} \right) , \qquad \mbox{where} \qquad  {\bf A}:= \left(\begin{array}{rr}
      -1 & \;\;\;1\;  \vspace{.2cm} \\  
      \;\;\;1 & -1 \; \\ 
   \end{array} \right) .
\eeqn

The basis of contours from Proposition \ref{prop_basis} in the space $\Lambda^*(z)$ with $z$ belonging to the part of a neighbourhood of $z=0$ lying in $\Pi^r$ consists of the contours $\calV_\ione,$ $\calV_\itwo$ counterclockwise encircling the points $\infty_\ione$ and $\infty_\itwo,$ respectively, and the contours $\calW_{\io;\ione},$ $\calW_{\io;\itwo}$ going from $\infty_\io$ to $\infty_\ione$ or $\infty_\itwo,$ respectively, such that inequality (\ref{gammaz1}) holds for their end-parts. These contours enter the basis set of contours in the given order: $\{\calV_\ione,\,\calV_\itwo,\,\calW_{\io;\ione},\,\calW_{\io;\itwo}\}.$  For our covering we may take: $\calV_\ione=\c^r_1-\c^r_2-\c^r_3+\c^r_4,\;$   $\calV_\itwo=\c^r_3-\c^r_4,\;$    $\calW_{\io;\ione} = \c^r_2,\;$ $\calW_{\io;\itwo}= \c^r_2+\c^r_4.$ 

The Jordan basis of contours consists of $\Gamma_\ione=3{\calV_\ione},$ $\Gamma_\itwo=-2\calW_{\io\ione}+\calW_{\io\itwo},$ $\Gamma_\ithree=3\calV_\itwo$ and $\Gamma_\ifour=-2\calW_{\io\itwo}+\calW_{\io\ione}.$
The connection  matrix (\ref{CM}) gives the coordinates of the Jordan basis with respect to the basis $\{\c^r_k\}.$ It has the form:
\beqs
C^{\io;\io,\io,\io} =    \left(\begin{array}{rrrr}
      3 & 0 & 0 & 0   \vspace{.2cm} \\  
      -3 & -1 & 0 & -1  \vspace{.2cm} \\  
      -3 & 0 & 3 & 0   \vspace{.2cm} \\  
      3 & 1 & -3 & -2  \\ 
   \end{array} \right) .
\eeqs
The contours $\calV_\ione,\,\calV_\itwo,\,\calW_{\io;\ione},\,\calW_{\io;\itwo}$ are the integration paths in formula (\ref{Psi0}) defining the solution $\tilde{\Psi}_\io(z)$ in a neighbourhood of $z=0.$ In the same way, the contours $\{\Gamma_k\}$ define the solution $\Psi_\io$ (\ref{Psi0J}) whose monodromy $M_{\io}$ at the origin is in the Jordan form. From the definition of the contours we find the monodromy matrix $\tilde{M}_\io$ of the solution $\tilde{\Psi}$ and its Jordan form $M_{\io}:$ 
\beqs
\tilde{M}_\io^{\io;\io,\io,\io} =    \left(\begin{array}{rrrr}
      -1 & 0 & -2 & -1   \vspace{.2cm} \\  
      0 & -1 & -1 & -2   \vspace{.2cm} \\  
      0 & 0 & -1 & 0   \vspace{.2cm} \\  
      0 & 0 & 0 & -1  \\ 
   \end{array} \right)  ; \hsp 
M_{\io}^{\io;\io,\io,\io} =    \left(\begin{array}{rrrr}
      -1 & 1 & 0 & 0   \vspace{.2cm} \\  
      0 & -1 & 0 & 0   \vspace{.2cm} \\  
      0 & 0 & -1 & 1   \vspace{.2cm} \\  
      0 & 0 & 0 & -1  \\ 
   \end{array}\right) .
   \eeqs

\subsection{The Hurwitz space $\M_{\ione;\io,\io}$}
\label{sect_H100}
The Hurwitz space $\M_{\ione;\io,\io}$ is the space of two-fold genus one coverings with four simple finite branch points $\l_1,\l_2,\l_3,\l_4.$ The corresponding Hurwitz diagram is shown in Figure \ref{fig_diagram100}. As before, we assume the branch cuts to be $[P_1,P_2]$ and $[P_3,P_4].$
\begin{figure}[htb]
\centering
\subfigure{\includegraphics[width=6cm]{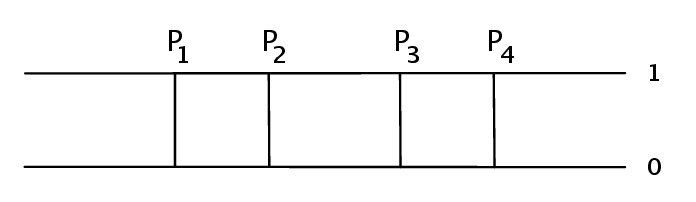}}
\caption{The Hurwitz diagram for the space $\M_{\ione;\io,\io}.$ }
\label{fig_diagram100}
\end{figure}

The coverings of this type can be obtained from the two-fold genus zero covering discussed in Section \ref{sect_H000} by adding one more branch cut to the existing sheets. We shall see the relationship between the Stokes matrices corresponding to the Hurwitz spaces  $\M_{\io;\io,\io}$ and $\M_{\ione;\io,\io}.$ This relationship gives a hint at how to proceed from the given Stokes matrix to the other if the corresponding coverings are obtained from each other by adding an extra branch cut (only if the sets of branch points on both coverings are ordered in the way shown in Figure \ref{fig_order}).

Figure \ref{fig_lr14} shows the contours $\c_k^r$ and $\c_k^l$ defining the solutions $\Psi^r(z)$ and $\Psi^l(z)$ (\ref{Psi_rl}) in the right and left half-planes, respectively.
\begin{figure}[htb]
\centering
\qquad\qquad\qquad{\includegraphics[width=4.5cm]{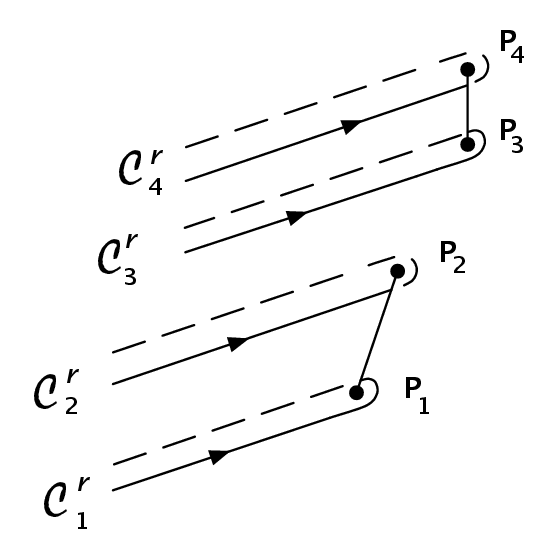}\qquad\subfigure{\includegraphics[width=7.5cm]{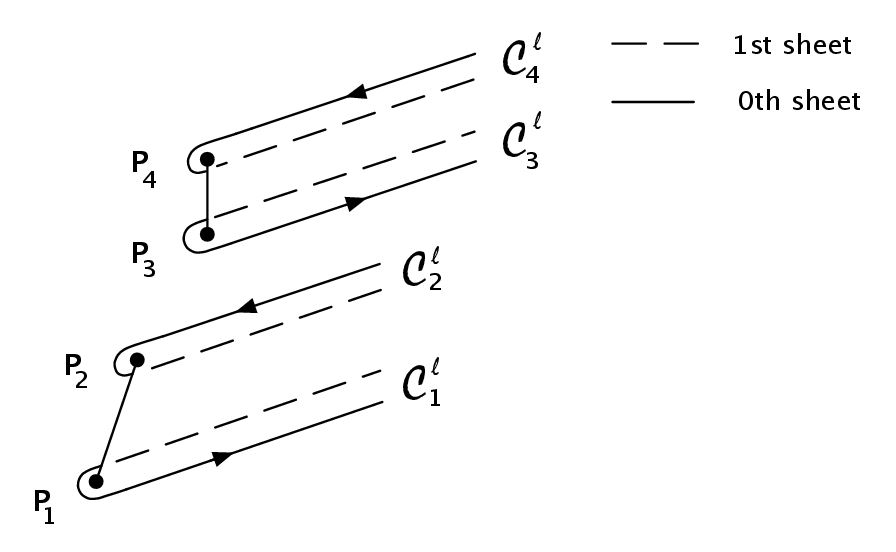}}}
\caption{Contours $\c_k^l$ and $\c_k^r$ for the space $\M_{\ione;\io,\io}.$}
\label{fig_lr14}
\end{figure}
 From Figure \ref{fig_lr14} we see that the deformed contours $\hat{\c}_k^r$ and $\hat{\c}_k^l$ in the space $\Lambda^*(z_+)$ are related as follows: 
\begin{equation*}
\hat{\c}_1^l=\hat{\c}_1^r - 2\hat{\c}_2^r +2\hat{\c}_3^r -2\hat{\c}_4^r ,\qquad \hat{\c}_2^l = \hat{\c}_2^r - 2\hat{\c}_3^r +2\hat{\c}_4^r,\qquad \hat{\c}_3^l=\hat{\c}_3^r-2\hat{\c}_4^r,\qquad \hat{\c}_1^l=\hat{\c}_4^r.
\end{equation*}
Thus, the Stokes matrix of the Frobenius manifold structure on the Hurwitz space $\M_{\ione;\io,\io}$ has the form: 
\beqn
\label{S100}
S^{\ione;\io,\io} =   \left(\begin{array}{cc}
      S^{\io;\io,\io} & \;\;{\bf 0}   \vspace{.2cm} \\  
     -2 {\bf A} & S^{\io;\io,\io}   \\  
   \end{array} \right) , 
\eeqn
where  $S^{\io;\io,\io}$  is the Stokes matrix (\ref{S000}) corresponding to the Hurwitz space $\M_{\io;\io,\io}$ and the $2\times 2$ matrix ${\bf A}$ is defined by (\ref{blockA}).

The basis of contours defining the solution $\tilde{\Psi}_\io$ in a neighbourhood of the origin consists of the contours $a$ and $b$, the contour $\calV_\ione$ encircling the point $\infty_\ione$ counterclockwise and the contour $\calW_{\io \ione}$ going from $\infty_\io$ to $\infty_\ione,$ see Proposition \ref{prop_basis}. We assume the contours enter the basis in the given order: $a,\,b,\,\calV_\ione,\,\calW_{\io \ione}.$ We may take: $a:=\c^r_1-\c^r_2,\;$ $b:=\c^r_2-\c^r_3,\;$ $\calV_\ione:=\c^r_1-\c^r_2+\c^r_3-\c^r_4,\;$ $\calW_{\io \ione}:=\c^r_4.$ 

The Jordan basis of contours is $\Gamma_\ione=a,$ $\Gamma_\itwo=b,$ $\Gamma_\ithree=2\calV_\ione$ and $\Gamma_\ifour=-\calW_{\io\ione}.$ The matrix of coefficients of these contours with respect to the basis $\{\c_k^r\}$ is the connection matrix $C^{\ione;\io,\io}$ given below. 
The monodromy matrix $\tilde{M}_\io^{\ione;\io,\io}$ of the solution  $\tilde{\Psi}_\io$ (\ref{Psi0}) is found 
using the definition of the contours $a,\,b,\,\calV_\ione,\,\calW_{\io \ione}.$
%
\beqs
C^{\ione;\io,\io}=   \left(\begin{array}{rrrr}
      1 & 0 & 2 & 0 \vspace{.2cm} \\  
      -1 & 1 & -2 & 0 \vspace{.2cm} \\ 
      0 & -1 &2 & 0 \vspace{.2cm} \\
      0 & 0 & -2 &-1
   \end{array} \right) ; \hsp \hsp
\tilde{M}_\io^{\ione;\io,\io}=   \left(\begin{array}{rrrr}
      -1 & 0 & 0 & 0 \vspace{.2cm} \\  
      0 & -1 & 0 & 0 \vspace{.2cm} \\ 
      0 & 0 &-1 & -2 \vspace{.2cm} \\
      0 & 0 & 0 &-1
   \end{array} \right).
\eeqs
 The Jordan form of the matrix $\tilde{M}_\io^{\ione;\io,\io}$ agrees with the one that can be found using Theorem \ref{thm_monodromy} and the structure of the contours $\{\Gamma_k\}.$

\subsection{Hurwitz spaces in arbitrary genus}

In the above examples we saw how the Stokes matrix corresponding to the space $\M_{\io;\io,\io}$ of two-fold genus zero coverings is related to the Stokes matrices corresponding to the Hurwitz spaces of coverings obtained from a two-fold genus zero covering by one of the following two operations: i) adding one sheet and one branch cut (Section \ref{sect_H0000}); ii) adding one branch cut (Section \ref{sect_H100}).   The former  operation does not change the genus of the covering and the latter increases the genus by one. Iterating these operations we can obtain an arbitrary covering with simple branch points which is not ramified over the point at infinity and for which  the branch cuts can be chosen to join two neighbouring ramification points.
Therefore, we can describe the Stokes matrix corresponding to the space of coverings of this type basing on the above examples. This is done in the next proposition. Recall that the branch cuts are assumed to be added in a way that the entire set of branch points is ordered as described by assumptions A1 and A2 in Section \ref{sect_assumptions}.

\begin{proposition}
Consider a Hurwitz space of $N$-fold genus $g$ coverings which have only simple branch points and are not ramified over the point at infinity. In this space, consider a neighbourhood of the covering which satisfies assumptions A1, A2 and A3 from Section \ref{sect_assumptions}. Namely: i) the branch points of the covering are ordered according to Figure \ref{fig_order} with respect to a line $l$ which is an admissible line for equation (\ref{RHz}) built from a Frobenius structure on the Hurwitz space; ii) for any odd $k,$ the ramification points $P_k$ and $P_{k+1}$ (and only these points) are connected by a branch cut; iii) the sheets of the covering are ordered and in a neighbourhood of the ramification point $P_k$ the ray ${\rm arg}\;  x_k = 3\pi/4 - \pi/2$ belongs to the lower of the two sheets glued together at $P_k.$ 
 
Then the Stokes matrix of the corresponding Frobenius manifold has the following structure. 
Let us take a branch cut $[P_k,P_l],$ $l=k+1,$ joining the sheets number $N_k$ and $N_l$ where $N_k<N_l$ and describe the corresponding $k$th and $l$th columns of the Stokes matrix $S=(S_{ij}).$  The elements $S_{ik}$ and $S_{il}$ for $i<k$ vanish.
The diagonal block, the intersection of the $k$th and $l$th columns and $k$th and $l$th rows, is given by the Stokes matrix $S^{\io;\io,\io} =   \left(\begin{array}{rr}
      1 & \;\;0  \vspace{.2cm} \\  
      -2 & \;\;1  \\ 
   \end{array} \right) $ from (\ref{S000}) corresponding to the Hurwitz space of two-fold genus zero coverings.
 For $i>l$ and $j=i+1$ the block
\beqn
\label{blockij}
\left(\begin{array}{rr}
      S_{ik} & S_{il}  \vspace{.2cm} \\  
      S_{jk} & S_{jl}\\ 
   \end{array} \right) 
\eeqn
is given by either ${\bf A}=\left(\begin{array}{rr}
      -1 & \;\;\;1\;  \vspace{.2cm} \\  
      \;\;\;1 & -1 \; \\ 
   \end{array} \right)$ from (\ref{blockA}), or $-{\bf A},$ or by $-2{\bf A},$ or vanishes. 

The block (\ref{blockij}) equals ${\bf A}$ in the following two cases.  
\begin{itemize}
\item The branch cut $[P_i,P_j]$ joins the sheets number $N_l$ and $N_j$ such that $N_l<N_j.$
\item The branch cut $[P_i,P_j]$ joins the sheets number $N_k$ and $N_i$ such that $N_i<N_k.$
\end{itemize}

The block (\ref{blockij}) equals $-{\bf A}$ if the inequalities between $N_l,$ $N_j$ and $N_k,$ $N_i$ are opposite to the above ones. 

The block (\ref{blockij}) equals $-2{\bf A}$ if the branch cut $[P_i,P_j]$ joins the sheets number $N_k$ and $N_l,$ i.e., the same sheets as the branch cut $[P_k,P_l].$

The block (\ref{blockij}) vanishes if the branch cuts $[P_i,P_j]$ and $[P_k,P_l]$ belong to four pairwise distinct sheets. 
\label{prop_Stokes}
\end{proposition}

{\it Proof.} 
The $k$th and $l$th columns of the Stokes matrix $S$ give the coordinates of the contours $\hat{\c}_k^l$ and $\hat{\c}_l^l$ with respect to the basis $\{\hat{\c}_j^r\}$ in the space $\Lambda^*(z_+)$ with $z_+\in l_+.$ 
The contour $\hat{\c}_k^l$  is obtained from the contour $\hat{\c}_k^r$ as follows. The projection $\l(\hat{\c}_k^r)$ of $\hat{\c}_k^r$ on the base of the covering is rotated  clockwise about the point $\l_k$ towards the projection $\l(\hat{\c}_k^l)$ and then is lifted back to the covering so that the resulting transformation of the contour $\hat{\c}_k^r$ is smooth in a neighbourhood of $P_k.$ Away from this neighbourhood the transformation is not smooth since 
the branch cuts belonging to the sheets number $N_k$ and $N_l$ are in the way of the rotation of the contour $\hat{\c}_k^r$ on the covering (see Figures \ref{fig_lr04}, \ref{fig_lr14} and the corresponding examples). Thus, the difference  $\hat{\c}_k^l-\hat{\c}_k^r$ can be expressed as a linear combination of contours encircling the branch cuts on the sheets $N_k$ and $N_l.$ As we have seen in the examples, a contour encircling a branch cut $[P_i,P_j]$ is given by 
$\pm(\hat{\c}_i^r-\hat{\c}_j^r)$ in the case of simple ramification. 

The assumptions A1 and A2 on the arrangements of the branch points and cuts imply that the branch cut $[P_i, P_j]$ can be in the way of the rotation of the contour $\hat{\c}_k^r$ towards $\hat{\c}_k^l$ on the covering only if $k<i<j.$ Therefore, in the $k$th column of the Stokes matrix the first $k-1$ elements vanish, i.e., $S_{ik}=0$ for $i<k.$ Analogously, for the $l$th column we have $S_{il}=0$ for $i<l.$ 

The clockwise (towards $\hat{\c}_l^l$) rotation of $\hat{\c}_l^r$  does not cross the branch cut $[P_k,P_l],$ therefore, $\hat{\c}_l^l$ is equivalent to $\hat{\c}_l^r$ plus the contours $\pm(\hat{\c}_i^r-\hat{\c}_j^r)$ with $l<i<j$ encircling the branch cuts
$[P_i,P_j]$ belonging to the sheets $N_k$ and/or $N_l.$ 
For the $k$th column of the Stokes matrix we note that the contour encircling the branch cut $[P_k,P_l]$ can be expressed as $\hat{\c}_k^r-\hat{\c}_l^r=\hat{\c}_k^l+\hat{\c}_l^l,$ therefore, $\hat{\c}_k^l = \hat{\c}_k^r-2\hat{\c}_l^r+\sum_{j>i>l} a_{ij}(\hat{\c}_i^r-\hat{\c}_j^r)$ with some integers $a_{ij},$ which shows that the diagonal block is given by the matrix $S^{\io;\io,\io}$ (\ref{S000}). 

The exact form of the added combinations of contours $\pm(\hat{\c}_i^r-\hat{\c}_j^r)$ with $l<i<j$ is analogous to that in the examples considered above. The form of the block (\ref{blockij}) given in the proposition is obtained by a straightforward generalization of the examples from Sections \ref{sect_H0000} and \ref{sect_H100}. 
Note that we need to specify which sheet lies above (is labeled with a larger number) because of the assumption that the contours of integration $\c^r_k$ in a neighbourhood of the corresponding ramification points pass from the lower to the upper sheet. 
$\Box$

To illustrate Proposition \ref{prop_Stokes} we give below the Stokes matrix $S$ corresponding to the Hurwitz space of the coverings of the type represented by the diagram in Figure \ref{fig_diagram_many} (note that the points $P_7$ and $P_8$ do not belong to the sheet number two).
\begin{figure}[htb]
\centering
\subfigure{\includegraphics[width=7cm]{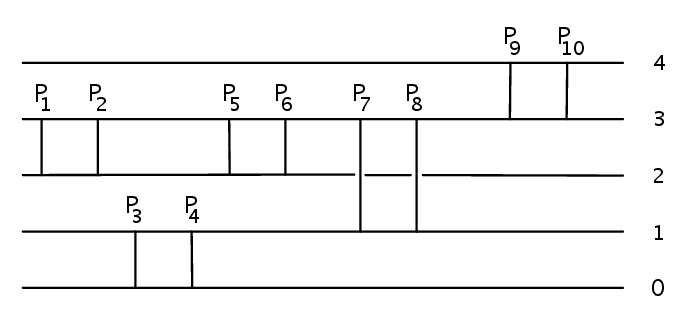}}
\caption{A Hurwitz diagram for the space $\M_{\ione;\io,\io,\io,\io,\io}.$ }
\label{fig_diagram_many}
\end{figure}
\beqs
S =   \left(\begin{array}{rcrcc}
       S^{\io;\io,\io}& {\bf 0} & {\bf 0} & {\bf 0} & {\bf 0} \vspace{.2cm} \\  
      {\bf 0} & S^{\io;\io,\io} & {\bf 0} & {\bf 0} & {\bf 0} \vspace{.2cm} \\ 
                  -2{\bf A} & {\bf 0} & S^{\io;\io,\io} & {\bf 0} & {\bf 0}  \vspace{.2cm} \\ 
                  -{\bf A} & {\bf A} & -{\bf A} & S^{\io;\io,\io} & {\bf 0} \vspace{.2cm} \\ 
                  {\bf A} & {\bf 0} & {\bf A} & {\bf A} & S^{\io;\io,\io}  \vspace{.2cm} \\ 
                                \end{array} \right),
\eeqs
where ${\bf 0}$ is a $2\times 2$ zero matrix; $S^{\io;\io,\io}$ is the Stokes matrix (\ref{S000}) corresponding to the Hurwitz space of genus zero two-fold coverings with two finite branch points, and ${\bf A}$ is the $2\times 2$ block given by (\ref{blockA}).

In particular, we get the Stokes matrix corresponding to the space of the genus zero coverings  obtained from the two-fold genus zero covering shown in Figure \ref{fig_diagram000} by consequently adding an extra sheet and two finite branch points. An example of such coverings is given by the Hurwitz diagram shown in Figure \ref{fig_diagram_stair}. 
\begin{figure}[htb]
\centering
\subfigure{\includegraphics[width=7cm]{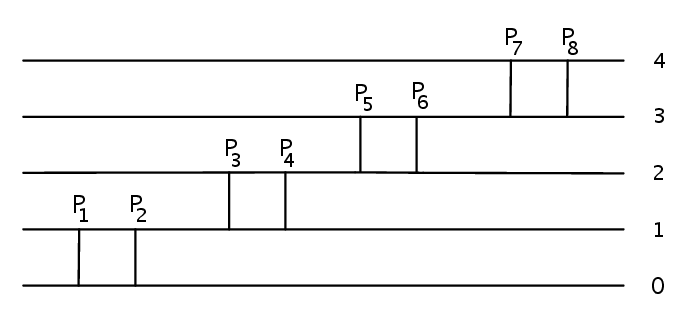}}
\caption{A Hurwitz diagram for the space $\M_{\io;\io,\io,\io,\io}.$ }
\label{fig_diagram_stair}
\end{figure}
In terms of the matrices $S^{\io;\io,\io}$ (\ref{S000}) and ${\bf A}$ (\ref{blockA}), the Stokes matrix for Frobenius manifold structures on the Hurwitz space of the coverings from Figure \ref{fig_diagram_stair} is given by: 
\beqs
S =   \left(\begin{array}{cccc}
       S^{\io;\io,\io}& {\bf 0} & {\bf 0} & {\bf 0}  \vspace{.2cm} \\  
      {\bf A} & S^{\io;\io,\io} & {\bf 0} & {\bf 0}  \vspace{.2cm} \\ 
                  {\bf 0} & {\bf A} & S^{\io;\io,\io} & {\bf 0}   \vspace{.2cm} \\ 
                  {\bf 0} & {\bf 0} & {\bf A} & S^{\io;\io,\io}  \vspace{.2cm} \\ 
                                \end{array} \right).
\eeqs
The Stokes matrix corresponding to the general case of Hurwitz space of such genus zero co\-ve\-rings without ramification over the point at infinity has the same structure.

In the case of the Hurwitz spaces $\M_{g;\io,\io}$ of hyperelliptic coverings represented by the Hurwitz diagram in Figure \ref{fig_diagram_hyperelliptic},
\begin{figure}[htb]
\centering
\subfigure{\includegraphics[width=7cm]{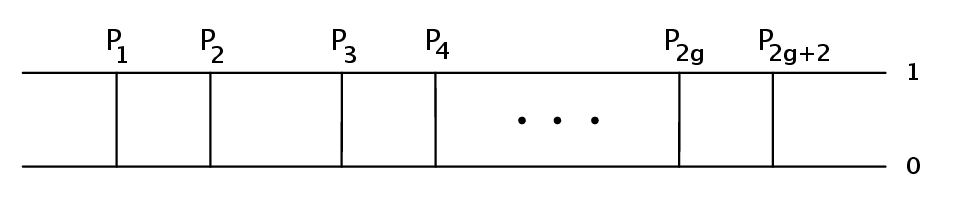}}
\caption{A Hurwitz diagram for the space $\M_{g;\io,\io}.$ }
\label{fig_diagram_hyperelliptic}
\end{figure}
Proposition \ref{prop_Stokes} implies that the $(2g+2) \times(2g+2) $ Stokes matrix of the corresponding Frobenius manifolds has the form:
\beqs
S =   \left(\begin{array}{ccccc}
       S^{\io;\io,\io}& {\bf 0} & {\bf 0} & {\bf 0} &{\dots} \vspace{.2cm} \\  
      -2{\bf A} & S^{\io;\io,\io} & {\bf 0} & {\bf 0}  &{\dots} \vspace{.2cm} \\ 
                 -2{\bf A} & -2{\bf A} & S^{\io;\io,\io} & {\bf 0} &{\dots}  \vspace{.2cm} \\ 
                  -2{\bf A} & -2{\bf A} & -2{\bf A} & S^{\io;\io,\io}  &{\dots} \vspace{.2cm} \\ 
                   \dots &\dots & \dots & \dots &{\dots} \vspace{.2cm} \\ 
                                \end{array} \right).
\eeqs

\vspace{.5cm}
{\bf Acknowledgments.} I am grateful to Arend Bayer, Claus Hertling, Dmitry Korotkin, Anton Mellit and Ilya Zakharevich for many valuable discussions and hints and to the Max-Planck-Institut f\"ur Mathematik for excellent working conditions.

\end{document}